\documentclass[12pt,aps,showpacs,notitlepage,longbibliography,floatfix]{revtex4-1}
\usepackage{amssymb}
\usepackage{amsmath}
\usepackage{bm}
\usepackage[caption=false]{subfig} 
\usepackage[pdftex]{graphicx}
\usepackage{multirow}
\usepackage{color}
\usepackage{courier}

\newcommand{\beq}{\begin{equation}}
\newcommand{\beqn}{\begin{eqnarray}}
\newcommand{\eeq}{\end{equation}}
\newcommand{\eeqn}{\end{eqnarray}}
\newcommand{\bfig}{\begin{figure}}
\newcommand{\efig}{\end{figure}}

\newcommand{\Hou}{ \ {\rm km \ s^{-1} \> Mpc^{-1}} }
\newcommand{\kms}{ \ {\rm km \ s^{-1} } }

\newcommand{\zehnum}{1.87} %1.75 WMAP9
\newcommand{\zhsnum}{1.48} % 1.43 WMAP9
\newcommand{\zindnum}{3.65} %3.728 WMAP9
\newcommand{\zindinfnum}{9.99} %9.58 WMAP9

\newcommand{\zindnumclosed}{2.38} %2.6 WMAP9
\newcommand{\zindnumopen}{5.25} %4.10, Planck, OL=0.65, 4.66 WMAP9, OL=0.65 

\newcommand{\chiindnum}{23.10}  % Ro Xind 
\newcommand{\zmaxnum}{8.55} % highest observed GRB host galaxy redshift

\newcommand{\chimaxflatnum}{30.31}  %30.44 flat WMAP9
\newcommand{\chimaxclosednum}{28.77}  %29.22 closed WMAP9
\newcommand{\chimaxopennum}{31.55}  %30.72 Planck, OL=0.65, 31.20 open WMAP9, OL=0.65

\newcommand{\vrecindnum}{1.86} % 1.86 in WMAP9, doesn't change much with Planck
\newcommand{\vpecznum}{0.1} %or 0.1 the z at which peculiar velocities dominate.  For planck cosmology

\newcommand{\Honum}{67.3} %69.32 WMAP9
\newcommand{\hnum}{0.673} %0.6932 WMAP9
\newcommand{\thnum}{14.53} %14.11 Gyr WMAP9
\newcommand{\OMnum}{0.315} % 0.2865 WMAP9
\newcommand{\OLnum}{0.685} % 0.7135 WMAP9
\newcommand{\ORADnum}{9.289 \times 10^{-5}} % 8.697 \times 10^{-5} WMAP9

\newcommand{\OKnum}{0}
\newcommand{\OTnum}{1}

\newcommand{\OLnumclosed}{0.800}
\newcommand{\OKnumclosed}{-0.115}
\newcommand{\OTnumclosed}{1.115}

\newcommand{\OLnumopen}{0.570} %0.650
\newcommand{\OKnumopen}{0.115} %0.035
\newcommand{\OTnumopen}{0.885} %0.965

\newcommand{\tauinfnum}{62.90} %62.56 Gyr WMAP9
\newcommand{\tauonum}{46.20} % 46.59 Gyr WMAP9
\newcommand{\taucmbnum}{0.91} %0.92 WMAP89

\newcommand{\tauonumdim}{3.18} % 3.303 WMAP9
\newcommand{\taucmbnumdim}{0.063} %0.0655 WMAP89

\newcommand{\chiAnum}{11.11} % 10.97 Glyr WMAP9
\newcommand{\chiBnum}{21.25} % 21.21 Glyr WMAP9
\newcommand{\chiABnum}{10.14} % 10.24 Glyr WMAP9

\newcommand{\tauAnum}{35.09} % 35.62 Gyr WMAP9
\newcommand{\tauBnum}{24.95} % 25.38 Gyr WMAP9
\newcommand{\tauABnum}{13.84} % 14.41 Gyr WMAP9

\newcommand{\alphafig}{135} % degrees for cones and int_plane figs

\newcommand{\zcmbnum}{1090.43} %1090.43 Planck
\newcommand{\alphacmbnum}{2.31} % Planck

\newcommand{\zbb}{$\infty$}
\newcommand{\zgal}{1.23} %1.27 WMAP9
\newcommand{\zearth}{0.41} % 0.42 WMAP9
 
\newcommand{\zeuk}{0.124}

\newcommand{\tlbb}{13.82} % 13.76 WMAP9
\newcommand{\tlgal}{8.80} % 8.81 Gyr lookback
\newcommand{\tlearth}{4.54} % 4.45 Gyr lookback
 
\newcommand{\tleuk}{1.65}

\newcommand{\tbb}{0}
\newcommand{\tgal}{5.01} % 4.96 WMAP9, 8.81 Gyr lookback
\newcommand{\tearth}{9.27} % 9.26 WMAP9, 4.45 Gyr lookback
 
\newcommand{\teuk}{12.16}

\newcommand{\Tabb}{0}
\newcommand{\Tabgal}{33.32} %33.58 WMAP9 8.81 Gyr lookback
\newcommand{\Tabearth}{40.81} % 41.32 WMAP9, 4.45 Gyr lookback
 
\newcommand{\Tabeuk}{44.45}

\newcommand{\zindbb}{3.65} % WMAP9 3.728
\newcommand{\zindgal}{0.506} % WMAP9 0.52
\newcommand{\zindearth}{0.195} % WMAP9 0.195
\newcommand{\zindeuk}{0.061}
\newcommand{\nameonea}{SDSS\_J031405.36-010403.8}
\newcommand{\nameoneb}{SDSS\_J171919.54+602241.0}
\newcommand{\nametwoa}{KX\_257}
\newcommand{\nametwob}{SDSS\_J110521.50+174634.1}
\newcommand{\namethreea}{Q\_0023-4124}
\newcommand{\namethreeb}{HS\_1103+6416}

\newcommand{\raonea}{48.5221}
\newcommand{\raoneb}{259.8313}
\newcommand{\ratwoa}{24.1229}
\newcommand{\ratwob}{166.3396}
\newcommand{\rathreea}{6.5496}
\newcommand{\rathreeb}{166.5446}

\newcommand{\deconea}{-1.0675}
\newcommand{\deconeb}{60.3781}
\newcommand{\dectwoa}{15.0481}
\newcommand{\dectwob}{17.7761}
\newcommand{\decthreea}{-41.1381}
\newcommand{\decthreeb}{64.0025}

\newcommand{\Rmagonea}{16.9}
\newcommand{\Rmagoneb}{18.6}
\newcommand{\Rmagtwoa}{16.7}
\newcommand{\Rmagtwob}{16.4}
\newcommand{\Rmagthreea}{14.2}
\newcommand{\Rmagthreeb}{14.7}

\newcommand{\Bmagonea}{20.1}
\newcommand{\Bmagoneb}{16.9}
\newcommand{\Bmagtwoa}{17.8}
\newcommand{\Bmagtwob}{25.1}
\newcommand{\Bmagthreea}{15.4}
\newcommand{\Bmagthreeb}{15.4}

\newcommand{\zaone}{6.109}
\newcommand{\zbone}{6.606}
\newcommand{\zatwo}{3.167}
\newcommand{\zbtwo}{6.086}
\newcommand{\zathree}{1.950}
\newcommand{\zbthree}{2.203}

\newcommand{\alpone}{116.003}
\newcommand{\alptwo}{130.355}
\newcommand{\alpthree}{154.357}

\newcommand{\colorone}{red}
\newcommand{\colortwo}{green}
\newcommand{\colorthree}{blue}

\newcommand{\onec}[1]{\textcolor{\colorone}{#1}}
\newcommand{\twoc}[1]{\textcolor{\colortwo}{#1}}
\newcommand{\threec}[1]{\textcolor{\colorthree}{#1}}

\newcommand{\mnras}{Mon. Not. Royal Astr. Soc.}

\newcommand{\pasa}{Pub. Astr. Soc. Australia}

\newcommand{\apjs}{Astrophys J. Suppl.}
\newcommand{\aap}{  Astron. \& Astrophys. }
\newcommand{\aj}{Astronomical J.}

\newcommand{\apjl}{Astrophys J. Lett.} %Tubbs & Wolfe

\begin{document}
%------------------------------------------------------------------------------------------------------------------------------------------------

\title{The Shared Causal Pasts and Futures of Cosmological Events}

\author{Andrew S. Friedman$^1$, David I. Kaiser$^1$, and Jason Gallicchio$^{2}$}

\email{Email addresses: asf@mit.edu; dikaiser@mit.edu; jason@frank.harvard.edu}

\affiliation{\\
$^1$Center for Theoretical Physics and Department of Physics, Massachusetts Institute of Technology, Cambridge, Massachusetts 02139 USA \\
$^2$Kavli Institute for Cosmological Physics, University of Chicago, Chicago, Illinois 60637 USA \\
}

\date{\today}

%------------------------------------------------------------------------------------------------------------------------------------------------

\begin{abstract} 
We derive criteria for whether two cosmological events can have a shared causal past or a shared causal future, assuming a Friedmann-Lemaitre-Robertson-Walker universe with best-fit $\Lambda$CDM cosmological parameters from the {\it Planck} satellite. We further derive criteria for whether either cosmic event could have been in past causal contact with our own worldline since the time of the hot ``big bang," which we take to be the end of early-universe inflation. 
We find that pairs of objects such as quasars on opposite sides of the sky with redshifts $z \geq \zindnum$ have no shared causal past with each other or with our past worldline. 
More complicated constraints apply if the objects are at different redshifts from each other or appear at some relative angle less than $180^\circ$, as seen from Earth. We present examples of observed quasar pairs that satisfy all, some, or none of the criteria for past causal independence. Given dark energy and the recent accelerated expansion, our observable universe has a finite conformal lifetime, and hence a cosmic event horizon at current redshift $z = \zehnum$. We thus constrain whether pairs of cosmic events can signal each other's worldlines before the end of time.  Lastly, we generalize the criteria for shared past and future causal domains for FLRW universes with nonzero spatial curvature.

\end{abstract}

\pacs{04.20.Gz; 98.80.-k; Preprint MIT-CTP 4440}

\maketitle

%-----------------------------------------------------------------------------------------------------------------------------------------------
\section{Introduction}

Universes (such as our own) that expand or contract over time can have nontrivial causal structure. Even in the absence of physical singularities, cosmic expansion can create horizons that separate observers from various objects or events \cite{rindler56,ellis93,davis04,faraoni11}. Our observable Universe has had a nontrivial expansion history: it likely underwent cosmic inflation during its earliest moments \cite{guth81,guth05,bassett06}; and observations strongly indicate that our Universe was decelerating after inflation and is presently undergoing a phase of accelerated expansion again, driven by dark energy consistent with a cosmological constant \cite{schmidt98,riess98,perlmutter99,astier06,woodvasey07,frieman08}. The late-time acceleration creates a cosmic event horizon that bounds the furthest distances observers will be able to see, even in infinite cosmic proper time \cite{starkman99,starobinsky00,loeb02}. 

One of the best-known examples of how nontrivial expansion history can affect causal structure concerns the cosmic microwave background radiation (CMB). At the time the CMB was emitted at redshift $z \approx 1090$ \cite{ade13}, too little time had elapsed since the hot big bang for regions on the sky separated by angles greater than about two degrees, as seen from the Earth today, to have exerted any causal influence on each other. The uniformity of the CMB temperature across the entire sky, including angles much greater than two degrees, is known as the ``horizon problem" \cite{dicke79,guth81,lineweaver05}. Early-universe inflation addresses the horizon problem by extending the past of our observable Universe to earlier times, prior to what is referred to as the hot ``big bang"; indeed, in this work, we will use the term ``big bang" to explicitly refer to the moment when early-universe inflation ends \cite{guth81,guth05,bassett06}. 

Modern astronomical observations have furnished huge datasets of distant objects at cosmologically interesting redshifts ($z \gtrsim \vpecznum$) with which we may explore causal structure beyond the example of the CMB (e.g. \cite{friedman05a,sakamoto11,paris12,flesch12,riess04b,riess07,kowalski08,hicken09a,amanullah10,rodney12}). We may ask, for example, whether two quasars that we observe today have been in causal contact with each other in the past. How far away do such objects need to be to have been out of causal contact between the hot big bang and the time they emitted the light we receive today?  Previous work investigating the uniformity of physical laws on cosmological scales has long emphasized the importance of observing causally disjoint quasars (e.g. \cite{tubbs80,pagel83}), culminating in recent searches for spatiotemporal variation of fundamental dimensionless constants such as the hydrogen fine-structure constant and the proton-to-electron mass ratio using quasar absorption lines (e.g. see \cite{king12a,king12b} and references therein, although see \cite{cameron12}). We add to such longstanding causal structure applications by outlining a novel formalism unifying past and future causal relations for cosmic event pairs, generalized for arbitrary space-time curvature, and applying it to the current best measurements of the cosmological parameters for our own Universe from the {\it Planck} satellite \cite{ade13}. An additional application, which will be explored in future work, centers on fundamental aspects of quantum mechanics where it is important to clarify whether physical systems are prepared independently on causal grounds alone.  Experiments designed along these lines might be able to test both fundamental physics and perhaps even specific models of inflation. This strongly motivates developing a secure handle on the theoretical conditions for past causal independence of cosmic event pairs, which is the primary focus of this work.  
 
In this paper we derive criteria for events to have a shared causal past --- that is, whether the past-directed lightcones from distant emission events overlap with each other or with our own worldline since the time of the big bang (at the end of inflation). If event pairs have no shared causal past with each other, no additional events could have jointly influenced both of them with any signals prior to the time they emitted the light that we observe today.  Similarly, if an event's past lightcone does not intersect our worldline, no events along Earth's comoving worldline could have influenced that event with any signals before the time of emission.  We find, for example, that objects like quasars on opposite sides of the sky with redshifts $z \geq \zindnum$ had been out of causal contact with each other and with our worldline between the big bang and the time they emitted the light we receive today. This critical value, which we call the causal-independence redshift, $z_{\rm ind} = \zindnum$, is not particularly large by present astronomical standards; tens of thousands of objects have been observed with redshifts $z > z_{\rm ind}$ (e.g. quasars from the Sloan Digital Sky Survey and other surveys \cite{paris12,flesch12}). More complicated past causal independence constraints apply if the objects are at different redshifts from each other or appear at some relative angle (as seen from Earth) less than $180^\circ$. The criteria depend on cosmological parameters such as the Hubble constant and the relative contributions to our Universe from matter, radiation, and dark energy. Using the current best-fit parameters for a spatially flat cosmology with dark energy and cold dark matter ($\Lambda$CDM), we derive conditions for past causal independence for pairs of cosmic objects at arbitrary redshift and angle. We also generalize these relationships for spacetimes with nonzero spatial curvature.

In addition to considering objects' shared causal pasts, we also investigate whether they will be able to exchange signals in the future, despite the late-time cosmic acceleration and the associated cosmic event horizon. By studying the overlap of objects' future lightcones with each other's worldlines, we determine under what conditions signals from various objects (including Earth) could ever reach other distant objects.  Our discussion of both the shared causal futures and causal pasts of cosmic event pairs is presented within a unified formalism.

Throughout the paper we assume that our observable Universe may be represented by a simply-connected, non-compact Friedmann-Lemaitre-Robertson-Walker (FLRW) metric, which is consistent with recent measurements of large-scale homogeneity and isotropy \cite{larson11,hinshaw12,PlanckIsotropy13}. In Section II we establish units and notation for distances, times, and redshifts. In Section III we derive the conditions required for past causal independence in the case of a spatially flat FLRW metric, and in Section IV we derive comparable relations for FLRW metrics of nonzero spatial curvature. Section V considers future lightcone intersections, and concluding remarks follow in Section VI. Appendix A revisits early-universe inflation and cosmic horizons within the formalism established in Sections II - III, and Appendix B examines the evolution of the ``Hubble sphere," beyond which objects recede from our worldline faster than light. 

\clearpage
%----------------------------------------------------------------------------------------------------------------------------------------------------------------------------------
\section{Distances, Times, and Redshifts}
\label{sec:dist_t_z}
%----------------------------------------------------------------------------------------------------------------------------------------------------------------------------------

For arbitrary spatial curvature, we may write the FLRW line-element in the form
\beq
ds^2 = - c^2 dt^2 + R_0^2 a^2(t) \left[ \frac{d\tilde{r}^2}{(1 - k \tilde{r}^2)} + \tilde{r}^2 \left( d\theta^2 + \sin^2\!\theta \ d \varphi^2 \right) \right] ,
\label{dsgeneralr}
\eeq
where $a(t)$ is the scale factor, $c$ is the speed of light, $R_0$ is a constant with units of length, and the dimensionless constant $k = 0, \pm 1$ indicates the curvature of spatial sections. (By including $R_0$, we take $a(t)$ and $\tilde{r}$ to be dimensionless for any spatial curvature $k$.) The angular coordinates range over $0 \leq \theta \leq \pi$ and $0 \leq \varphi \leq 2\pi$, and in the case $k = 1$, the radial coordinate satisfies $\tilde{r} \leq 1$. We normalize $a(t_0) = 1$, where $t_0$ is the present time.

For arbitrary curvature $k$, the (dimensionless) comoving radial distance $\chi$ between an object at coordinate $\tilde{r}$ and the origin is given by
\beq
\chi = \int_{0}^{\tilde{r}}  \frac{d \tilde{r}'}{\sqrt{1 - k \tilde{r}'^2} } =   \left\{ \begin{array}{cc}
\arcsin \> \tilde{r}    &\>\> {\rm for} \>\> k = 1 , \\
\tilde{r}                  &\>\> {\rm for} \>\> k = 0 , \\
{\rm arcsinh} \> \tilde{r}  &\>\> {\rm for} \>\> k = -1 .
\end{array} \right.
\label{chidef}
\eeq
We may likewise define a (dimensionless) conformal time, $\tau$, via the relation
%%%%%%%%%%
\beq
d\tau \equiv  \frac{c}{R_0} \frac{dt}{a(t)} .
\label{taudef}
\eeq
Then we may rewrite the line-element of Eq. (\ref{dsgeneralr}) as
%%%%%%%%%
\beq
ds^2 = R_0^2 a^2 (\tau) \left[ - d\tau^2 + d\chi^2 + S_k^2 (\chi) \left( d \theta^2 + \sin^2 \theta d\varphi^2 \right) \right] ,
\label{dsgeneralchi}
\eeq
where
%%%%%%%%%%
\beq
S_k (\chi) = \left\{ \begin{array}{cc}
\sin \chi               &\>\> {\rm for} \>\> k = 1 , \\
\chi                      &\>\> {\rm for} \>\> k = 0 , \\
{\rm sinh}\> \chi &\>\> {\rm for} \>\> k = -1 .
\end{array} \right.
\label{Sk}
\eeq
It is also convenient to define
\beq
C_k (\chi) \equiv \sqrt{1- k S_k^2(\chi) } = \left\{ \begin{array}{cc}
\cos \chi     &\>\> {\rm for} \>\> k = 1 , \\
1              &\>\> {\rm for} \>\> k = 0 , \\
\cosh \chi  &\>\> {\rm for} \>\> k = -1 .
\end{array} \right.
\label{Ck}
\eeq
Given Eq. (\ref{dsgeneralchi}), light rays traveling along radial null geodesics $(d \theta=d\varphi=0)$ obey
%%%%%%%%%
\beq
d\chi =  d\tau .
\label{nullgeochi}
\eeq

For any spatial curvature $k$, we set the dimensionful constant $R_0$, with units of length, to be
%%%%%%%%%%
\beq
R_0 = \frac{c}{H_0} ,
\label{R0}
\eeq
where $H_0$ is the present value of the Hubble constant with best-fit value $H_0 =  \Honum \Hou = (\thnum \>\> {\rm Gyr})^{-1}$ \cite{ade13}. 
%{\bf [DK: update all numerical terms here in light of Planck satellite data; the main one to change is the present best-fit value of $H_0$.]} 
In the case $k = 1$, the coordinates $(\tilde{r}, \theta, \varphi)$ only cover half the spatial manifold. In that case, $\tilde{r} = \sin (0) = 0$ at the north pole and $\tilde{r} = \sin (\pi/2) = 1$ at the equator, so for a single-valued radial coordinate $\tilde{r}$, we may only cover the upper (or lower) half of the manifold. We may avoid this problem by working with the coordinate $\chi$ in the $k = 1$ case and allowing $\chi$ to range between $0 \leq \chi \leq \pi$ rather than $0 \leq \chi \leq \pi / 2$ \cite{peebles93,peacock99}. 

The cosmological redshift, $z$, of an object whose light was emitted at some time $t_e$ and which we observe today at $t_0$ is given by
%%%%%%%%
\beq
1 + z = \frac{a (t_0 )}{a (t_e)} = \frac{1}{a_e} ,
\label{redshiftdef}
\eeq
upon using our normalization convention $a(t_0) = 1$ and defining $a_e \equiv a(t_e)$. Following \cite{peebles93,hogg99}, we parameterize the Friedmann equation governing the evolution of $a(t)$ in terms of the function
%%%%%%%
\beq
E(a) \equiv \frac{H (a)}{H_0} = \sqrt{ \Omega_\Lambda + \Omega_k a^{-2} + \Omega_M a^{-3} + \Omega_R a^{-4} } ,
\label{Edef}
\eeq
where $H(a)$ is the Hubble parameter for a given scale factor $a=a(t)$. The $\Omega_i$ are the ratios of the energy densities contributed by dark energy ($\Omega_\Lambda$), cold matter ($\Omega_M$), and radiation ($\Omega_R$) to the critical density $\rho_c = 3H_0^2/(8 \pi G)$, where $G$ is Newton's gravitational constant.  We also define a fractional density associated with spatial curvature ($\Omega_k \equiv 1 - \Omega_{\rm T}$) and the total fractional density of matter, dark energy, and radiation ($\Omega_{\rm T} \equiv \Omega_M + \Omega_\Lambda + \Omega_R$). We assume that $\Omega_\Lambda$ arises from a genuine cosmological constant with equation of state $w = p / \rho = -1$, which is consistent with recent measurements \cite{larson11,hinshaw12,riess11,woodvasey07,astier06,ade13}, and hence $\Omega_\Lambda a^{-3 (1 + w)} = \Omega_\Lambda$. Current observations yield best-fit cosmological parameters for our Universe consistent with 
%%%%%%%
\beq
\begin{split}
\vec{\Omega} = (h, \Omega_M, \Omega_\Lambda, \Omega_R, \Omega_k,  \Omega_{\rm T}) = (\hnum, \OMnum, \OLnum, \ORADnum, \OKnum, \OTnum) ,
\end{split}
\label{Omegavalues}
\eeq
where we define the dimensionless Hubble constant as $h \equiv H_0 / (100 \Hou)$.  Values for Eq. (\ref{Omegavalues}) are taken from Table 2, column 6 of \cite{ade13} including the most recent CMB temperature data from the {\it Planck} satellite and low multipole polarization data from the 9-year Wilkinson Microwave Anisotropy Probe (WMAP) release \cite{bennett12}.
%the rightmost column of Table 4 of \cite{hinshaw12} (labeled +eCMB+BAO+$H_0$) which includes joint cosmological constraints from the 9-year Wilkinson Microwave Anisotropy Probe (WMAP) data, other CMB data from the Atacama Cosmology Telescope \cite{das11,vanengelen12}, the South Pole Telescope \cite{keisler11}, measurements of Baryon Acoustic Oscillations \cite{blake11,beutler11,padmanabhan12,anderson12}, and the best independent estimates of $H_0$ \cite{riess11}.  
%{\bf [DK: again, update refs and main text to refer to Planck data.]} 
The fractional radiation density $\Omega_R$ is derived from the relation $\Omega_R = \Omega_M /(1+ z_{eq})$ where $\Omega_M = \Omega_b + \Omega_c$ is the fractional matter density given by the sum of the fractional baryon ($\Omega_b$) and cold dark matter ($\Omega_c$) densities and $z_{eq}$ is the redshift of matter-radiation equality. 
The quantities $\Omega_b h^2$, $\Omega_c h^2$, and $z_{eq}$ are all listed in Table 2, column 6 of \cite{ade13}.
%The quantities $\Omega_b$, $\Omega_c$, and $z_{eq}$ are all listed in Table 4 of \cite{hinshaw12}.

Given Eqs. (\ref{taudef}), (\ref{nullgeochi}), (\ref{redshiftdef}), and cosmological parameters from Eq. (\ref{Omegavalues}), we may evaluate comoving distance along a (radial) null geodesic using either $a(t)$ or $z$ as our time-like variable,
%%%%%%%%%
\beq
\chi =  \int_{a_e}^{1} \frac{da}{a^2 E(a)} = \int_{0}^{z} \frac{dz'}{E (z')} .
\label{chiz}
\eeq
Although Eq. (\ref{chiz}) does not permit analytic solutions for the general case in which the various $\Omega_i$ are nonvanishing, the equation may be integrated numerically to relate comoving distance to redshift. 

We may also consider how conformal time, $\tau$, evolves. If $\tau = 0$ is the beginning of time and inflation did not occur, $\tau$ is equivalent to the comoving distance to the particle horizon,
\beqn
\label{eq:tau}
\tau (t) =  \int_{0}^{a_{e}} \frac{da}{a^2 E(a)} =  \int_{z}^{\infty} \frac{dz'}{E(z')} .
\eeqn
As above, $\tau$ is dimensionless and $R_0 \tau / c = H_0^{-1} \tau$ has dimensions of time. The present age of the Universe, $\tau_0 = \tau (t_0)$, is given by
\beqn
\label{eq:tau_o}
\tau_0 \equiv  \int_{0}^{1} \frac{da}{a^2 E(a)} =  \int_{0}^{\infty} \frac{dz}{E(z)} \equiv \chi_\infty
\eeqn
which is equivalent to $\chi_\infty$, the comoving distance to the particle horizon today (at the comoving location corresponding to $z=\infty$). 

Even if inflation did occur, Eq. (\ref{eq:tau}) is still a reliable way to calculate $\tau$ numerically for times after inflation, $\tau > 0$. We consider inflation to begin at some early cosmic time $t_i$ and to persist until some time $t_{\rm end}$, where $t_{\rm end}$ will typically be of the order $t_{\rm end} \sim {\cal O}(10^{-37} \> {\rm sec} )$ \cite{guth05,bassett06}. In this case, the limits of integration in Eq. (\ref{eq:tau}) would be altered as 
\beqn
\label{eq:tau_inflation}
\tau (t) =  \int_{a(t_{\rm end})}^{a_{e}} \frac{da}{a^2 E(a)} =  \int_{z}^{z(t_{\rm end})} \frac{dz'}{E(z')} ,
\eeqn
where $a(t_{\rm end})$ is the scale factor at the end of inflation ($\tau (t_{\rm end}) = 0$) and $z (t_{\rm end})$ is the redshift for a hypothetical object we could observe today that emitted light at $\tau = 0$. Although $a (t)$ would have grown enormously during inflation, such that $a (t_{\rm end}) \gg a (t_i)$, we still expect $a (t_{\rm end}) \ll a_e$ for objects whose light was emitted well after the end of inflation. In particular, as discussed in Appendix A, for cosmological parameters as in Eq. (\ref{Omegavalues}) we have $a (t_{\rm end}) / a (t_0) \sim {\cal O} (10^{-28})$, so that the nonzero lower bound to the scale-factor integral in Eq. (\ref{eq:tau_inflation}) makes a negligible numerical contribution to the evolution of $\tau$ for $\tau > 0$ after the end of inflation. The same is true for the large but finite upper limit $z (t_{\rm end}) \sim {\cal O} (10^{28})$ in the integral over redshift in Eq. (\ref{eq:tau_inflation}). Thus we may still use Eq. (\ref{eq:tau}) to evaluate $\tau$ numerically for times after the end of inflation.

If inflation did occur, it would correspond to times $\tau < 0$. For convenience we assume $k = 0$ for the explicit construction, though comparable results may be derived for $k = \pm 1$ as well. Assuming quasi-de Sitter expansion during inflation, Eq. (\ref{taudef}) may be solved as
%%%%%%%%%%
\beq
\tau (t) =  \frac{1}{ a (t_{\rm end} ) } \left(  \frac{H_0}{ H_I}  \right) \left[ 1 - \frac{a (t_{\rm end} )}{a(t) } \right] ,
\label{tauinflation}
\eeq
where $H_I$ is the value of the Hubble parameter during inflation, and we have used Eq. (\ref{R0}) for $R_0$. As usual, we find that $\tau < 0$ during inflation, and $\tau \rightarrow 0^-$ as $t \rightarrow t_{\rm end}$. If we assume instant reheating to a radiation-dominated universe at $t_{\rm end}$, then we may match smoothly to a solution in which $\tau > 0$ following the end of inflation. In particular, for a radiation-dominated phase in a spatially flat FLRW universe we may write
%%%%%%%%%
\beq
a (t) = a (t_{\rm end}) \left( \frac{t}{t_{\rm end} } \right)^{1/2} 
\label{aRD}
\eeq
or
%%%%%%%%
\beq
\tau (t) =  \frac{2 H_0 t_{\rm end} }{ a(t_{\rm end} ) } \left[ \left( \frac{t}{t_{\rm end} } \right)^{1/2} - 1 \right] 
\label{tauRD}
\eeq
for $t \geq t_{\rm end}$. Consistent with Eqs. (\ref{tauinflation}) and (\ref{tauRD}), we therefore take the time of the big bang to be $t_{\rm end}$ or $\tau (t_{\rm end}) = 0$, after the end of early-universe inflation.

%----------------------------------------------------------------------------------------------------------------------------------------------------------------------------------
\section{Spatially Flat Case}
\label{sec:flat}
%----------------------------------------------------------------------------------------------------------------------------------------------------------------------------------

In this section we consider a spatially flat universe (like our own), and set $k = \Omega_k = 0$. We may then absorb the constant $R_0$ into the definition of the comoving radial coordinate by introducing $r \equiv R_0 \tilde{r}  = R_0 \chi$. For the remainder of this section, we work in terms of a comoving radial coordinate $r$ that carries dimensions of length, whereas the comoving radial coordinate $\chi$ remains dimensionless, as does conformal time $\tau$. In this section, boldface symbols represent spatial 3-vectors.
 
With respect to the CMB dipole, we treat the Earth's position in the CMB rest frame as the origin of the spatial coordinates.  However, small corrections between the heliocentric and CMB frame or systematic redshift offsets from peculiar velocities do not affect our results, which are presented only to 2 decimal places in redshift.  Typical random peculiar velocities of $\sigma_v^{\rm pec} \approx 300 \kms$ lead to a systematic redshift error of only $\sigma_z^{\rm pec} \approx 0.001$ \cite{davis11}.

We now present the formalism for intersection of past lightcones for cosmic event pairs in a flat universe (see Fig. \ref{Conformal}). An object A at comoving spatial location ${\bf r}_A$ emits light at conformal time $\tau_A$ which the observer on Earth receives at the present time, $\tau_0$, while an object B at comoving location ${\bf r}_B$ emits light at conformal time $\tau_B$ which the observer also receives at $\tau_0$. The light signals travel along null geodesics, $ds = 0$, and hence from Eq. (\ref{nullgeochi}) we immediately find
%%%%%%%%
\beq
\begin{split}
\tau_0 - \tau_A &= \chi_A = R_0^{-1} \vert {\bf r}_A \vert , \\
\tau_0 - \tau_B &= \chi_B = R_0^{-1} \vert {\bf r}_B \vert  .
\end{split}
\label{tauAtauB}
\eeq
The past-directed lightcones from the emission events A and B intersect at comoving location ${\bf r}_{AB}$ at time $\tau_{AB}$, such that
%%%%%%%%
\beq
\begin{split}
\tau_A - \tau_{AB} &= R_0^{-1} \vert {\bf r}_A - {\bf r}_{AB} \vert , \\
\tau_B - \tau_{AB} &= R_0^{-1} \vert {\bf r}_B - {\bf r}_{AB} \vert ,
\end{split}
\label{tABrAB}
\eeq
or, upon making use of Eq. (\ref{tauAtauB}),
%%%%%%%%%%%
\beq
\begin{split}
\tau_0 - \tau_{AB}  &= \chi_A + R_0^{-1} \vert {\bf r}_A - {\bf r}_{AB} \vert , \\
\tau_0 - \tau_{AB}  &= \chi_B + R_0^{-1} \vert {\bf r}_B - {\bf r}_{AB} \vert .
\end{split}
\label{tABrAB2}
\eeq
%
%%%%%%%%%%%%%%%%%%%%%%%%%%%%%%%%%%%%%%%%%%%%%%%%%%%%%%%%%%%%%%%%%
\bfig 
\centering
\includegraphics[width=6.5in]{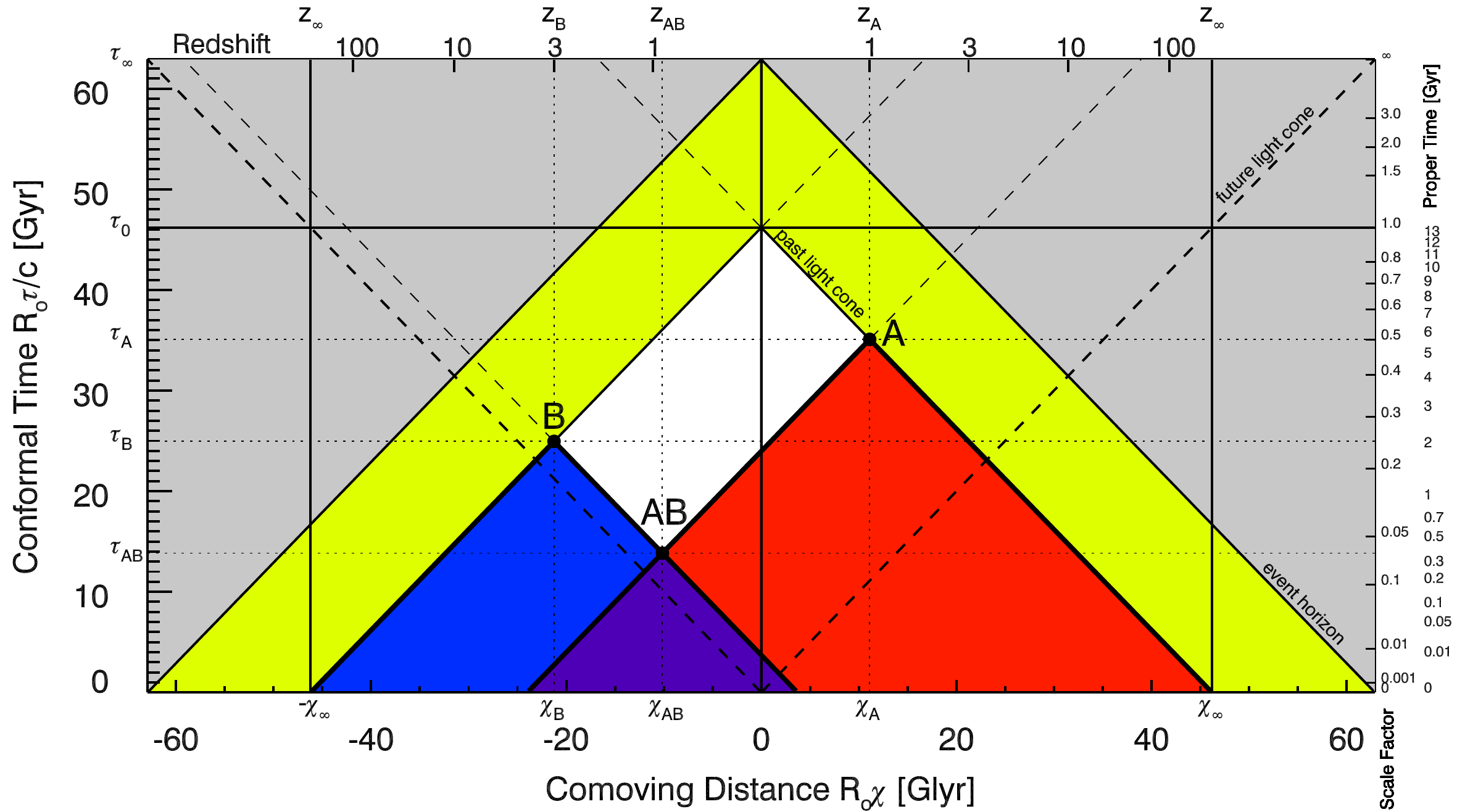}
%{hub_davis1_100_tau_1p0000000_3p0000000_180p00000_69p3_0p2865_0p71343303_8p6970249e-05_0p0000000.pdf}
\caption{\small \baselineskip 12pt 
Conformal diagram showing comoving distance, $R_0 \chi$ in Glyr, versus conformal time, $R_0 \tau / c$ in Gyr, for the case in which events A and B appear on opposite sides of the sky as seen from Earth ($\alpha = 180^\circ$). The observer sits at Earth at $\chi = 0$ at the present conformal time $\tau=\tau_0$. Light is emitted from A at $(\chi_A, \tau_A)$ and from B at $(\chi_B, \tau_B)$; both signals reach the Earth along our past lightcone at (0,$\tau_0$). The past-directed lightcones from the emission events (red and blue for A and B, respectively) intersect at $(\chi_{AB}, \tau_{AB})$ and overlap for $0< \tau < \tau_{AB}$ (purple region). For redshifts $z_A = 1$ and $z_B = 3$ and a flat $\Lambda$CDM cosmology with parameters given in Eq. (\ref{Omegavalues}), the events are located at comoving distances $R_0 \chi_A = \chiAnum$ Glyr and $R_0 \chi_B = \chiBnum$ Glyr, with emission at conformal times $R_0 \tau_A / c = \tauAnum$ Gyr and $R_0 \tau_B / c = \tauBnum$ Gyr. The past lightcones intersect at event AB at $R_0 \chi_{AB} = \chiABnum$ Glyr at time $R_0 \tau_{AB} / c = \tauABnum$ Gyr, while the present time is $R_0 \tau_0 / c = \tauonum$ Gyr. Also shown are the cosmic event horizon (line separating yellow and gray regions) and the future-directed lightcones from events A and B (thin dashed lines) and from the origin (0,0) (thick dashed lines). In a $\Lambda$CDM cosmology like ours, events in the yellow region outside our current past lightcone are space-like separated from us today but will be observable in the future, while events in the gray region outside the event horizon are space-like separated from observers on Earth forever. Additional scales show redshift (top horizontal axis) and time as measured by the scale factor, $a(\tau)$, and by proper time, $t$, (right vertical axis) as measured by an observer at rest at a fixed comoving location. 
}
\label{Conformal}
\efig

Without loss of generality, we consider event A to occur later than event B ($\tau_A > \tau_B$ and hence $z_A < z_B$), in which case the past-directed lightcone centered on A must expand further before it intersects with the past-directed lightcone centered on B. By construction, we take event B to lie along the $x$ axis and the vector ${\bf r}_A$ to make an angle $\theta$ with respect to the $x$ axis, so that an observer on Earth would see events A and B separated by an angle $\alpha = \pi - \theta$ on the sky. See Fig. \ref{Plane}.

%%%%%%%%%%%%%%%%%%%%%%%%%%%%%%%%%%%%%%%%%%%%%%%%%%%%%%%%%%%%%%%%%
\bfig
\centering
\includegraphics[width=3.21in]{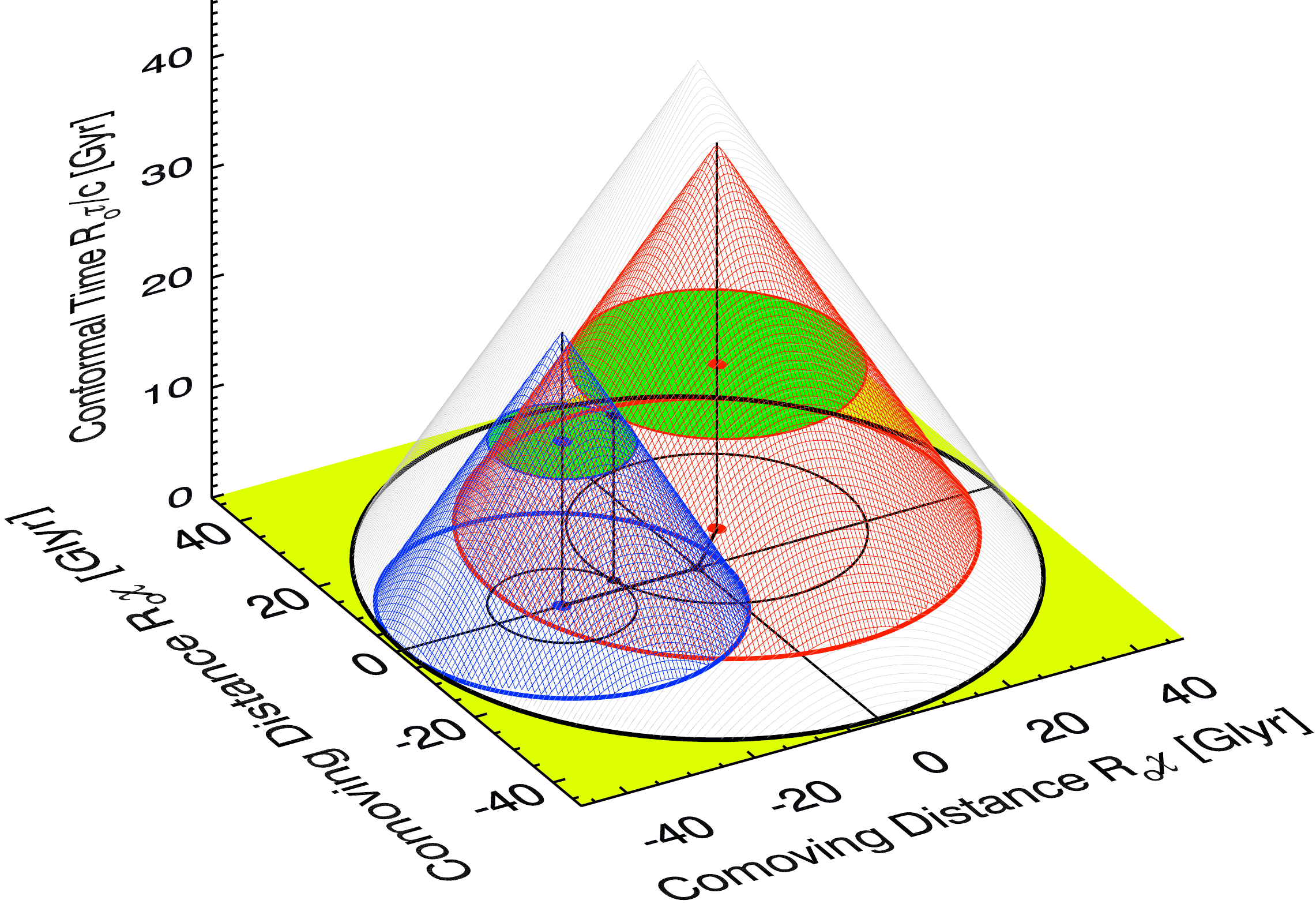}
%{cones_150_1p0000000_3p0000000_150p00000_69p3_0p2865_0p71343303_8p6970249e-05_0p0000000.pdf}
\includegraphics[width=3.21in]{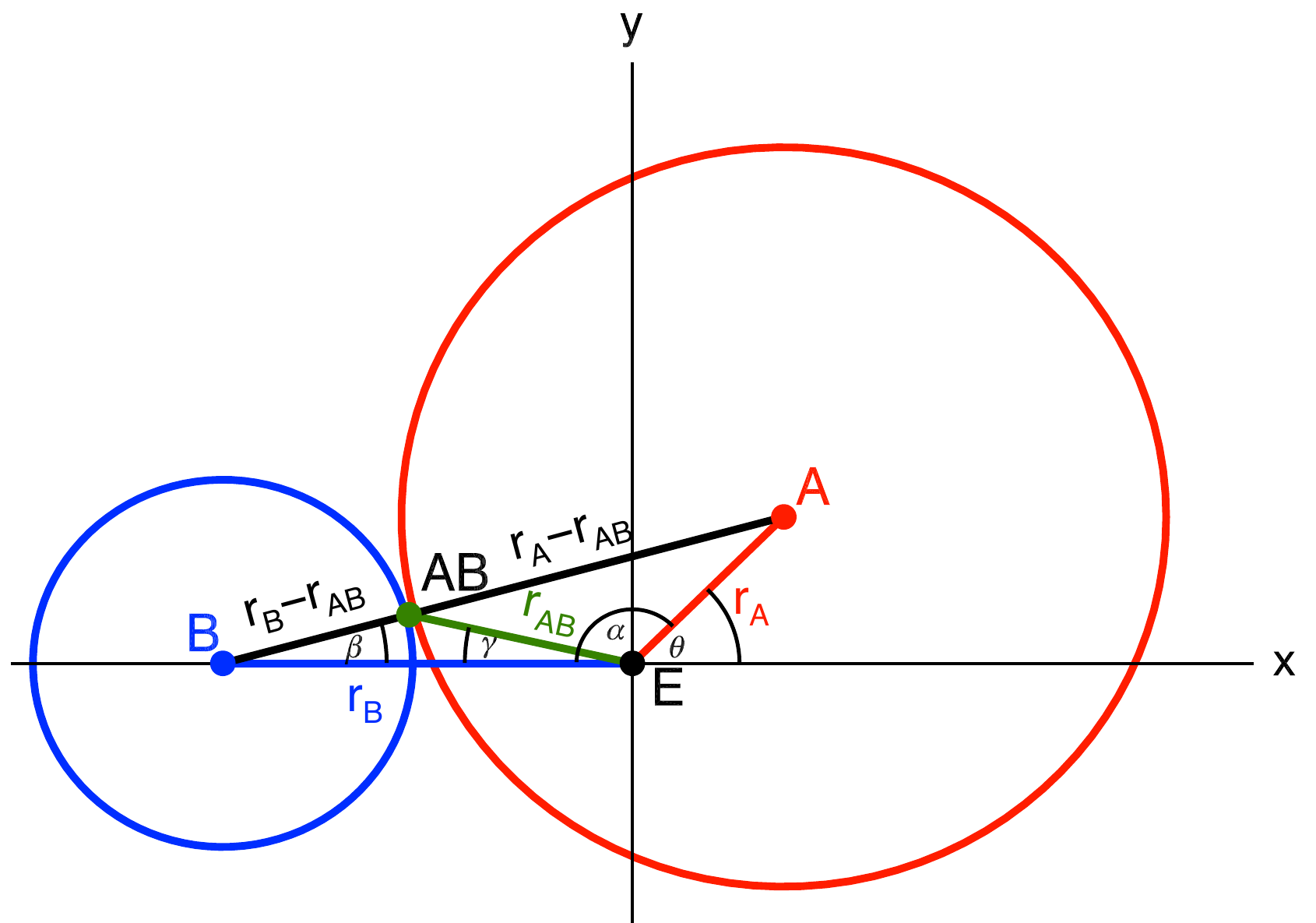}
\caption{\small \baselineskip 12pt 
({\it Left}) Plot of our past lightcone from $\tau_0$ (gray outer cone) and the past lightcones from emission events A and B (red and blue cones, respectively). The green circles show the projection of the past lightcones on the hypersurface $\tau=\tau_{AB}$ when the lightcones first intersect. For the case shown here, $\alpha = \alphafig^\circ$, $z_A = 1$, and $z_B = 3$. ({\it Right}) Plot of the spatial $(x,y)$ plane for the hypersurface $\tau=\tau_{AB}$, corresponding to the green circles in the left figure. Earth is at the origin. Event A occurs at comoving location ${\bf r}_A$ (red vector) and event B occurs at comoving location ${\bf r}_B$ (blue vector). The past-directed lightcones from A and B appear in the plane as circles centered on A and B, respectively. The past lightcones intersect at event AB at comoving location ${\bf r}_{AB}$ (green vector). The angle between events A and B as seen from Earth is $\alpha = \pi - \theta$. For animations of the intersecting lightcones as one varies $z_i$ and $\alpha$, see the online Supplementary Materials at \texttt{http://prd.aps.org/supplemental/PRD/v88/i4/e044038}, which include 11 animations with captions based on Figures 1 and 2, constructed for cosmological parameters from the {\it Planck} satellite given by Eq. 11. The animations vary or hold fixed the redshifts and angular separations of cosmic event pairs to illustrate the conditions for events to have either a shared causal past or no shared causal past since the big bang. These and other animations are also available at \texttt{http://web.mit.edu/asf/www/causal\_past.shtml}.
}
\label{Plane}
\efig
%%%%%%%%%%%%%%%%%%%%%%%%%%%%%%%%%%%%%%%%%%%%%%%%%%%%%%%%%%%%%%%%%

Given the orientation of the vectors in Fig. \ref{Plane}b, we have 
%%%%%%%%%
\beq
\vert {\bf r}_A - {\bf r}_B \vert = \vert {\bf r}_A - {\bf r}_{AB} \vert + \vert {\bf r}_B - {\bf r}_{AB} \vert .
\label{rArBmagnitude}
\eeq
Using Eqs. (\ref{tABrAB}) and (\ref{rArBmagnitude}), we then find
%%%%%%%%%
\beq
\tau_{AB} = \frac{1}{2} \left( \tau_A + \tau_B - \chi_L \right)  ,
\label{tauABgeneral}
\eeq
where we have defined $\chi_L$ as the (dimensionless) comoving spatial distance between events A and B:
%%%%%%%
\beq
\begin{split}
\chi_L &\equiv R_0^{-1} \vert {\bf r}_A - {\bf r}_B \vert \\
&= R_0^{-1} \sqrt{ ( {\bf r}_A - {\bf r}_B ) \cdot ( {\bf r}_A - {\bf r}_B ) }  \\
&= \sqrt{ \chi_A^2 + \chi_B^2 - 2 \chi_A \chi_B \cos \alpha  } \> .
\end{split}
\label{rL}
\eeq
In the special case $\alpha = \pi$ (see Fig. 1), for which $\chi_L \rightarrow \chi_A + \chi_B$, Eq. (\ref{tauABgeneral}) reduces to
%%%%%%%%%%%%%
\beq
\tau_{AB} \rightarrow  \tau_A + \tau_B - \tau_0 
\eeq
upon using Eq. (\ref{tauAtauB}).

We may also solve for the comoving spatial location, ${\bf r}_{AB}$, at which the past-directed lightcones intersect. Squaring both sides of the identity ${\bf r}_A = {\bf r}_B + ({\bf r}_A - {\bf r}_B )$ yields
%%%%%%%%%%
\beq
r_A^2 = r_B^2 + r_L^2 - 2 r_B r_L \cos \beta ,
\label{rA1}
\eeq
where $\beta$ is the angle between vectors ${\bf r}_B$ and $({\bf r}_B - {\bf r}_A)$, as in Fig. \ref{Plane}b, and $r_L = \vert {\bf r}_A - {\bf r}_B \vert = R_0 \chi_L$. We likewise have
%%%%%%%%%
\beq
{\bf r}_{AB} \cdot {\bf r}_{AB} = \left[ {\bf r}_B - \left( {\bf r}_B - {\bf r}_{AB} \right) \right] \cdot \left[ {\bf r}_B - \left( {\bf r}_B - {\bf r}_{AB} \right) \right] .
\label{rAB1a}
\eeq
Upon using $r_{AB} = R_0 \chi_{AB}$ and Eq. (\ref{tABrAB}) to substitute $\vert {\bf r}_B - {\bf r}_{AB} \vert = R_0 (\tau_B - \tau_{AB} )$, Eq. (\ref{rAB1a}) may be written
%%%%%%%%%%
\beq
\chi_{AB}^2 = \chi_B^2 -2 \chi_B (\tau_B - \tau_{AB} ) \cos \beta + (\tau_B - \tau_{AB} )^2 .
\label{rAB1}
\eeq
From Eqs. (\ref{rA1}) and (\ref{rAB1}), we then find
%%%%%%%%
\beq
\chi_{AB}^2 = \chi_B^2 + \left( \tau_B - \tau_{AB} \right)^2 - \frac{2 \chi_B}{\chi_L} \left( \tau_B - \tau_{AB} \right) \left( \chi_B - \chi_A \cos \alpha \right) .
\label{rABfinal}
\eeq

By fixing $\alpha$ and $\chi_B$ and using Eqs. (\ref{tauAtauB}), (\ref{tauABgeneral}), and (\ref{rL}), we may derive the condition on the critical comoving distance $\hat{\chi}_A$ such that the past lightcones from A and B intersect at time $\tau_{AB}$,
\beqn
\label{eq:XaTab_flat}
\hat{\chi}_A = \frac{  \chi_B - ( \tau_0 - \tau_{AB} ) }{ \left[  \frac{ \chi_B (1 + \cos \alpha )}{  2 (\tau_0 - \tau_{AB} ) } - 1 \right] } .
\eeqn
Alternatively, we may fix $\chi_A$ and $\chi_B$ to derive the crititcal angle $\hat{\alpha}$ such that the past lightcones intersect at $\tau_{AB}$,
\beqn
\label{eq:alphaTab_flat}
\hat{\alpha}  =  \cos^{-1}\left( \frac{\chi_A^2 + \chi_B^2 - (\tau_A + \tau_B -2\tau_{AB})^2}{2 \chi_A \chi_B }\right) .
\eeqn
When $\tau_{AB} \le 0$, events A and B share no causal past after the end of inflation.  Considering event pairs that just barely meet this condition ($\tau_{AB} = 0$) leads to Figs. \ref{ABindep} and \ref{za_theta}, where we use Eq. (\ref{eq:XaTab_flat}) with $\tau_{AB} = 0$ to plot the hyperbolic curves for different angles $\alpha$ in Fig. \ref{ABindep}a and Fig. \ref{za_theta}. For Fig. \ref{ABindep}b, we must invert Eq. (\ref{chiz}) numerically to solve for the redshift $z$ corresponding to a given comoving distance $\chi (z)$. Setting $\tau_{AB} = 0$, then for $\chi_A \ge \hat{\chi}_{A}$ or $\alpha \ge \hat{\alpha}$, events A and B share no causal past since the big bang. In particular, if we fix $\alpha = \pi$ and consider the symmetric case in which $\chi_A = \chi_B$, then Eq. (\ref{eq:XaTab_flat}) for $\tau_{AB} = 0$ and cosmological parameters $\vec{\Omega}$ as in Eq. (\ref{Omegavalues}) yields $R_0 \chi_{\rm ind} = \chiindnum$ Glyr, which, using Eq. (\ref{chiz}), corresponds to the causal-independence redshift $z_{\rm ind} = \zindnum$.

%%%%%%%%%%%%%%%%%%%%%%%%%%%%%%%%%%%%%%%%%%%%%%%%%%%%%%%%%%%%%%%%%
\bfig
\centering
\includegraphics[width=3.2in]{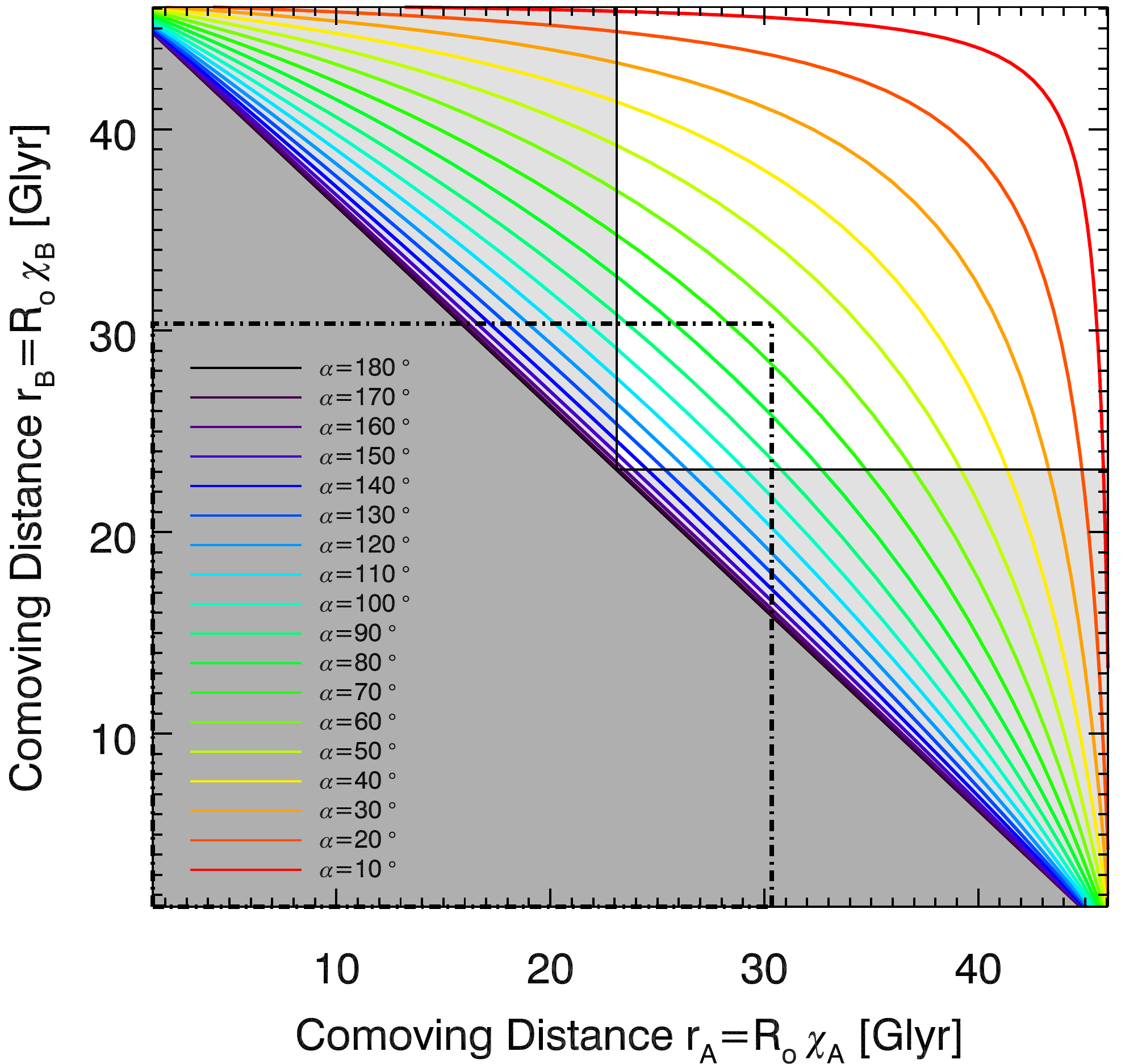}
\includegraphics[width=3.2in]{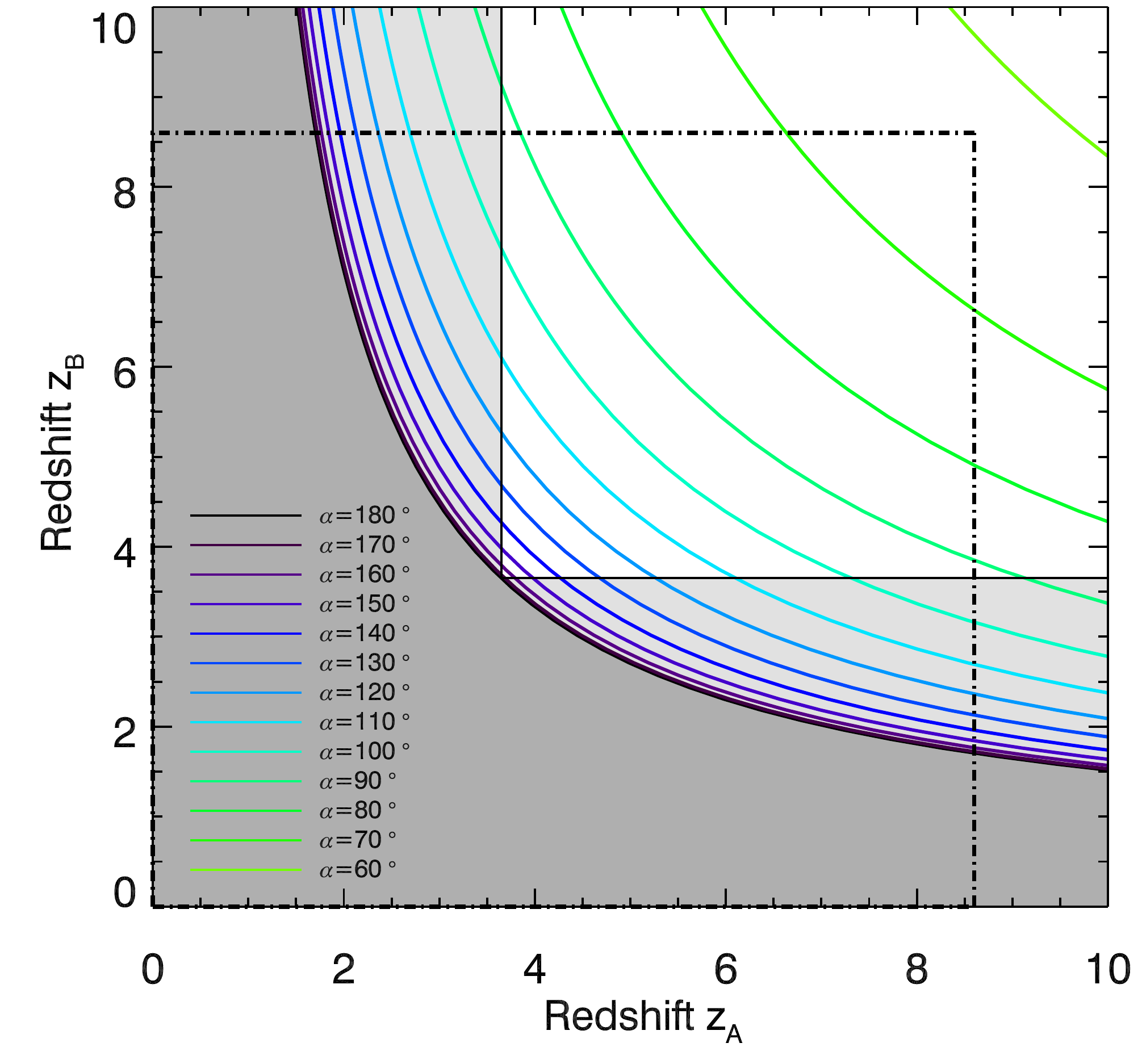}
%{lc_intersect_BB_za_zb_50_12_0_10_60p000000_180_sym_box_69p3_0p2865_0p71343303_8p6970249e-05_0p0000000.pdf} 

\caption{\small \baselineskip 12pt  
({\it Left}) Comoving distance $R_0 \chi_A$ versus $R_0 \chi_B$ for pairs of objects separated by angle $\alpha$, such that ({\it a}) their past-directed lightcones intersect at $\tau_{AB} = 0$ (colored curves for various angles), and ({\it b}) neither object's past-directed lightcone intersects our worldline after $\tau = 0$ (white box in upper right corner). For a given $\alpha$, comoving distances for event pairs that lie above the corresponding colored curve (toward the upper right corner) satisfy $\tau_{AB} < 0$ and thus share no causal connection after the end of inflation.  Event pairs with comoving distances in the light gray region have at least one object with a past lightcone that intersects our worldline at some time $\tau > 0$; thus the Earth's comoving location had been in causal contact with the event prior to emission. Objects in the lower left of the plot (dark gray region) have $\tau_{AB} > 0$ and hence always have a shared causal past for any angular separation. For $\alpha = 180^{\circ}$ and $\chi_A = \chi_B$, objects with $R_0 \chi > R_0 \chi_{\rm ind} = \chiindnum \> {\rm Glyr}$ share no causal past with each other or with our worldine since $\tau = 0$. ({\it Right}) The same plot in terms of redshift rather than comoving distance. For $\alpha = 180^\circ$ and $z_A = z_B$, object pairs with $z > z_{\rm ind} = \zindnum$ share no causal past with each other or with our worldline since $\tau = 0$. Both plots are constructed for a flat $\Lambda$CDM cosmology with parameters $\vec{\Omega}$ given in Eq. (\ref{Omegavalues}). In both figures, the dashed black box corresponds to the most distant object observed to date, at $z_{\rm max} = \zmaxnum$ or $R_0 \chi_{\rm max} = \chimaxflatnum \> {\rm Glyr}$, corresponding to the Gamma-Ray Burst in associated host galaxy UDFy-38135539 \cite{lehnert10}. 
}
\label{ABindep}
\efig
%%%%%%%%%%%%%%%%%%%%%%%%%%%%%%%%%%%%%%%%%%%%%%%%%%%%%%%%%%%%%%%%%
\bfig
\centering
\includegraphics[width=3.5in]{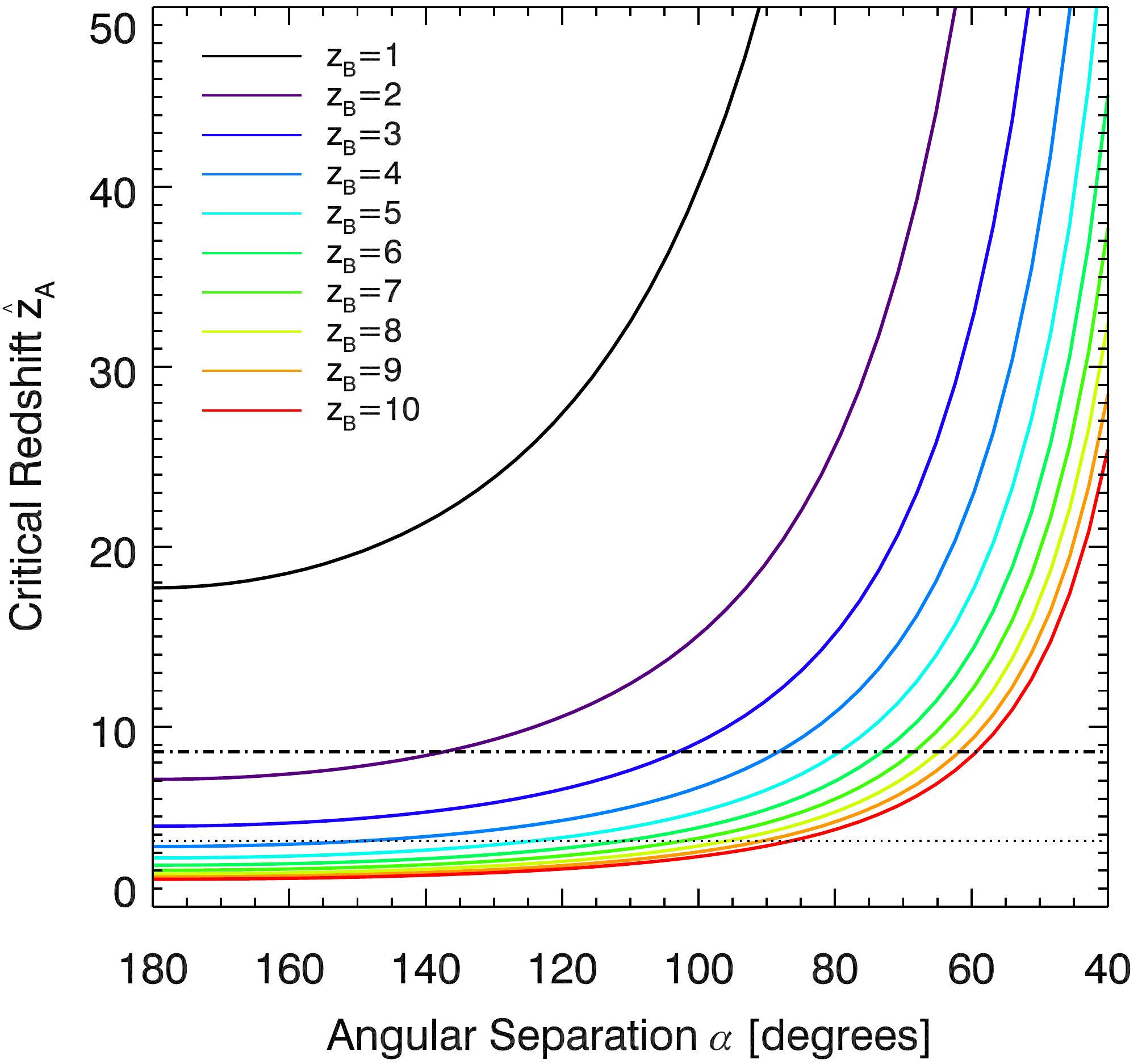} 
\caption{\small \baselineskip 12pt
For various fixed values of $z_B$, we plot the critical redshift $\hat{z}_A$ vs. the angular separation $\alpha$ such that $\tau_{AB}=0$. For each $z_B$ and $\alpha$, $\hat{z}_A$ is derived from $\hat{\chi}_{A}$ in Eq. (\ref{eq:XaTab_flat}) by inverting Eq. (\ref{chiz}) numerically. For all values of $z_B$, $\hat{z}_A$ monotonically increases as $\alpha$ decreases:  as the angular separation between event pairs decreases, larger redshifts for object A (for a given $z_B$) are required for the events to have no shared causal past. Event pairs with $z_A > \hat{z}_A$ that lie above the colored curve for a given $\alpha$ and $z_B$ have no shared causal past since the end of inflation.  
For any angle $\alpha \le 180^{\circ}$, events A and B have no shared causal past with Earth's worldline if $z_A > z_{\rm ind}=\zindnum$ (above the thin dashed line) and $z_B > z_{\rm ind}=\zindnum$. As in Fig. \ref{ABindep} the dashed horizontal line corresponds to the most distant object observed to date, at $z_{\rm max} = \zmaxnum$. 
}
\label{za_theta}
\efig
%%%%%%%%%%%%%%%%%%%%%%%%%%%%%%%%%%%%%%%%%%%%%%%%%%%%%%%%%%%%%%%%%

We may further impose the condition that neither event A nor B shares a causal past with our own worldline since $\tau = 0$. From Eq. (\ref{nullgeochi}), for $\tau \geq 0$ the comoving distance to the future-directed lightcone emanating from the origin $(\chi,\tau)=(0,0)$ is given by
%%%%%%%%
\beq
\chi_{\rm flc} (\tau) = \tau .
\label{dhor}
\eeq
See Fig. \ref{Conformal}. If inflation did not occur and $\tau = 0$ corresponds to $t = 0$, then $\chi_{\rm flc}(\tau) = \chi_{\rm ph} (\tau)$, the comoving distance to the particle horizon for an observer at rest at $\chi=0$. 
%The future-directed lightcone from the origin $(0,0)$ will intersect with the past-directed lightcone originating from Earth today (0,$\tau_0$), at some location $\chi_{\rm lc}$ at conformal time $\tau_{\rm lc}$. 
Along the radial null geodesic extending backward from Earth at $(\chi,\tau)=(0,\tau_0)$ toward the event at A, the past-directed lightcone is given by
%%%%%%%%%
\beq
\chi_{\rm plc} (\tau)  =  \tau_0 - \tau .
\label{xbackward}
\eeq
The past-directed lightcone from (0,$\tau_0$) will intersect the future-directed lightcone from $(0,0)$ at some location $\chi_{\rm lc}$ at conformal time $\tau_{\rm lc}$
%%%%%%%%%
\beq
\chi_{\rm plc} (\tau_{\rm lc} ) = \chi_{\rm flc}(\tau_{\rm lc}) 
\eeq
or
%%%%%%%%
\beq
\tau_{\rm lc} = \frac{1}{2} \tau_0 .
\label{taulc}
\eeq
As long as $\tau_A < \tau_{\rm lc} = \tau_0 / 2$, 
%then the emission event at A will occur outside the future-directed lightcone of the observer's worldline (whose origin is at the big bang at $\tau = 0$). 
then the past lightcone from event A will not intersect the observer's worldline since the big bang at $\tau = 0$.
By construction, since we have identified $\tau_A \geq \tau_B$, the past lightcone of event B will likewise not intersect the observer's worldline since $\tau = 0$.
%emission event at B likewise will occur outside the future-directed lightcone of the origin. 
For $\vec{\Omega}$ as in Eq. (\ref{Omegavalues}), the requirement that $\tau_A < \tau_0 / 2$ is satisfied by any object with $z_A > z_{\rm ind} = \zindnum$. See Fig.~\ref{ABindep}.

Requiring {\it both} $\tau_{AB} \leq 0$ {\it and} $\tau_B \leq \tau_A < \tau_0 / 2$ ensures that events A and B share no causal past with each other and that neither shares any causal past with our own worldline since the time of the big bang at $\tau = 0$. A quick examination of Fig. \ref{Conformal} illustrates that if the emission events A and B have no shared causal past with each other or with us since $\tau = 0$, then neither will any prior events along the worldlines of A and B. Many real objects visible in the sky today fulfill the conditions $\tau_{AB} \leq 0$ and $\tau_B \leq \tau_A < \tau_0 / 2$. Representative astronomical objects (quasar pairs) that obey all, some, or none of these joint conditions are displayed in Fig. \ref{ABempirical} and listed in Table \ref{QTable}.

Of course, one may consider objects that have been out of causal contact with each other only during more recent times. For example, one may calculate the criteria for objects' past lightcones to have shared no overlap since the time of the formation of the thin disk of the Milky Way galaxy around $\tlgal$ Gyr ago \cite{Peloso05}; or since the formation of the Earth $\tlearth$ Gyr ago \cite{Dalrymple01}; or since the first appearance on Earth of eukaryotic cells (precursors to multicellular organisms) $\tleuk$ Gyr ago \cite{knoll06}. Events more recent than around $1.35$ Gyr ago correspond to redshifts $z \leq 0.1$, and hence to distances where peculiar velocities are not negligible compared to cosmic expansion \cite{davis11}. For the $\alpha=180^{\circ}$ case, pushing the past-lightcone intersection time closer to the present day, $\tau_{AB} \rightarrow \tau_0$, yields curves in the $z_A$-$z_B$ plane that move down and to the left through the gray region of Fig. \ref{ABindep}b. See Fig. \ref{Tabindep} and Table II. 

%%%%%%%%%%%%%%%%%%%%%%%%%%%%%%%%%%%%%%%%%%%%%%%%%%%%%%%%%%%%%%%%%
\bfig
\centering
\includegraphics[width=3.5in]{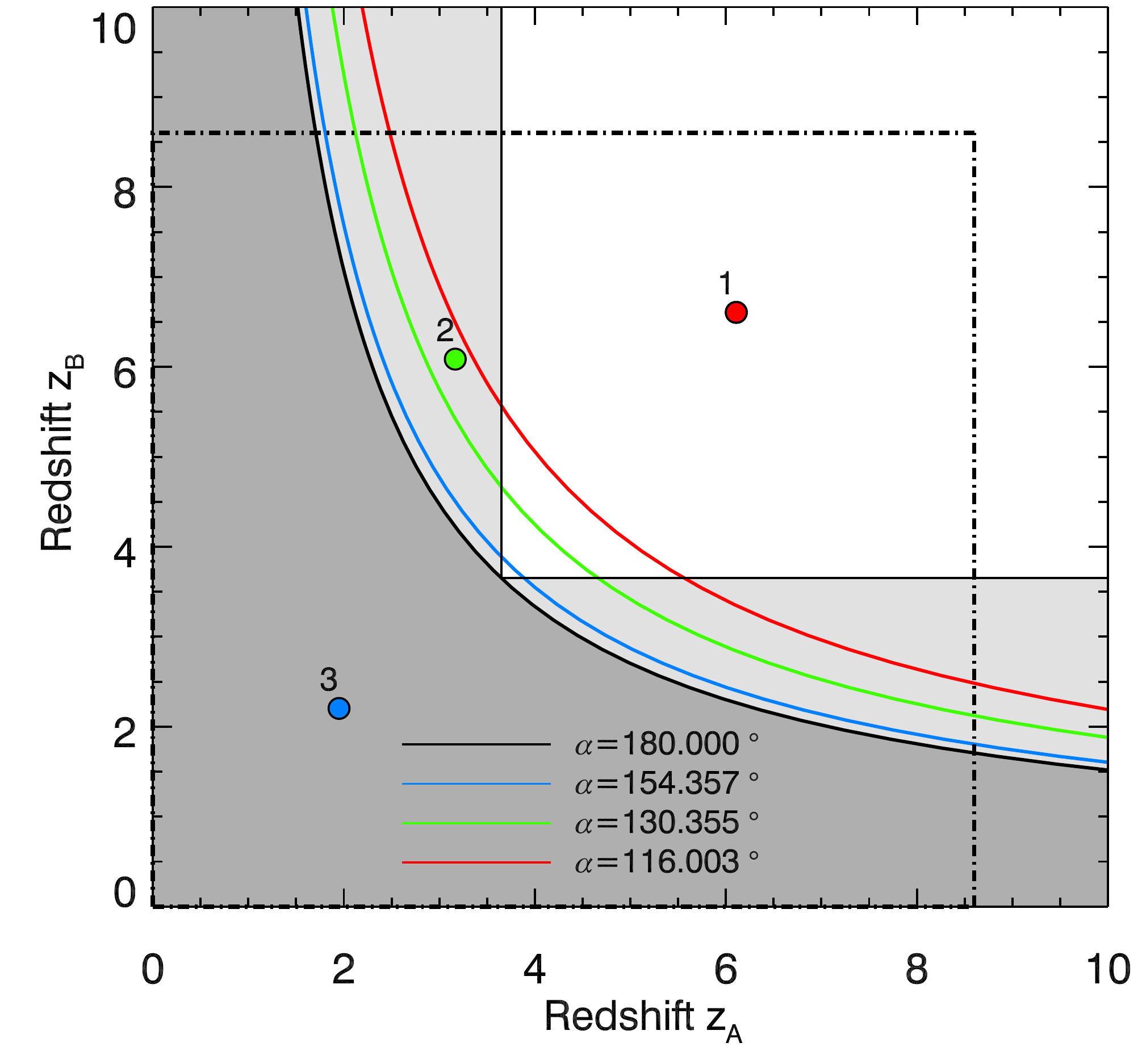}
\caption{\small \baselineskip 12pt 
Same as Fig. \ref{ABindep}b, with three quasar pairs marked (see Table~\ref{QTable}). For pair 1 (red), the past lightcones from each emission event share no overlap with each other or with our worldline since $\tau = 0$. For pair 2 (green), the past lightcones from each emission event share no overlap with each other, though the past lightcone from quasar $A_2$ does overlap our worldline for $\tau > 0$. For pair 3 (blue), both emission events have past lightcones that intersect each other as well as our worldline at times $\tau > 0$. 
 }
\label{ABempirical}
\efig
%%%%%%%%%%%%%%%%%%%%%%%%%%%%%%%%%%%%%%%%%%%%%%%%%%%%%%%%%%%%%%%%%

\begin{table}
\begin{center}
{\scriptsize
\begin{tabular}{|c|c|c|c|c|c|c|c|c|}
\hline
{\bf Pair }                             & {\bf Separation}                                        & {\bf Event }            & {\bf Redshifts }                 & {\bf Object }        & {\bf RA }    & {\bf DEC }  & {\bf R}                  & {\bf B} \\ [-2ex]
                                                & {\bf Angle ${\alpha}_i$ [deg] }     & {\bf Labels }          & {${z_A}_i$, ${z_B}_i$ }         & {\bf Names }         & [deg]           & [deg]           &  [mag]                  & [mag] \\        
\hline
\onec{ \multirow{2}{*}{1} }    & \multirow{2}{*}{$\alpone$}                         & $A_1$                    & $\zaone$                         & \nameonea         & \raonea     & \deconea    & $\Rmagonea$  & $\Bmagonea$   \\
                                                &                                                                    & $B_1$                    & $\zbone$                         & \nameoneb        & \raoneb     & \deconeb    & $\Rmagoneb$  & $\Bmagoneb$   \\
\hline
\twoc{ \multirow{2}{*}{2} }    & \multirow{2}{*}{$\alptwo$}                          & $A_2$                    & $\zatwo$                         & \nametwoa         & \ratwoa     & \dectwoa     & $\Rmagtwoa$  & $\Bmagtwoa$   \\
                                               &                                                                      & $B_2$                   & $\zbtwo$                          & \nametwob          & \ratwob    & \dectwob      & $\Rmagtwob$  & $\Bmagtwob$   \\
\hline
\threec{ \multirow{2}{*}{3} }  & \multirow{2}{*}{$\alpthree$}                      & $A_3$                     & $\zathree$                      & \namethreea     & \rathreea   & \decthreea   & $\Rmagthreea$  & $\Bmagthreea$   \\
                                                &                                                                    & $B_3$                     & $\zbthree$                      & \namethreeb     & \rathreeb   & \decthreeb   & $\Rmagthreeb$  & $\Bmagthreeb$   \\

\hline
\end{tabular}
}
\caption{\small \baselineskip 12pt 
Three quasar pairs from \cite{flesch12}, as shown in Fig. \ref{ABempirical}. Redshift pairs (${z_A}_i$, ${z_B}_i$) and angular separations ${\alpha}_i$ (in degrees) are chosen so that the pairs obey all (pair 1), some (pair 2), or none (pair 3) of the joint conditions of having no shared causal past with each other ($\tau_{AB} \leq 0$) and each having no shared causal past with our worldline ($\tau_A , \tau_B < \tau_0 / 2$). Given the parameters in Eq. (\ref{Omegavalues}), the latter constraint corresponds to $z_A, z_B > \zindnum$. Basic properties of each quasar from \cite{flesch12} are also shown including: object names from the relevant quasar catalogs, celestial coordinates $(RA,DEC)$ in degrees, and $R$ and $B$ band brightnesses (in magnitudes). }

\label{QTable}

\end{center}
\vspace{-0.6cm}
\end{table}

%%%%%%%%%%%%%%%%%%%%%%%%%%%%%%%%%%%%%%%%%%%%%%%%%%%%%%%%%%%%%%%%%
\bfig
\centering
\includegraphics[width=3.2in]{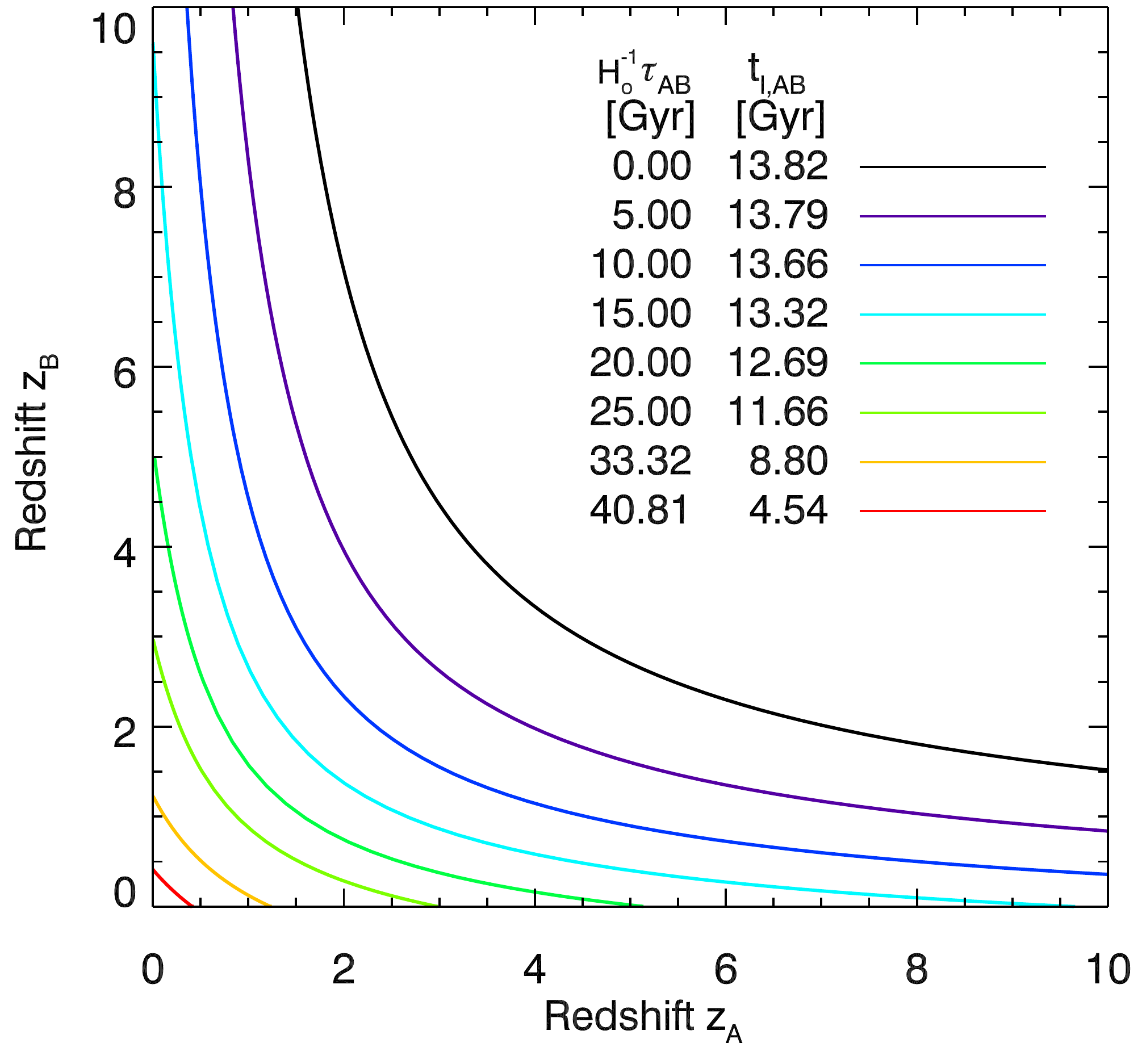} 
\includegraphics[width=3.2in]{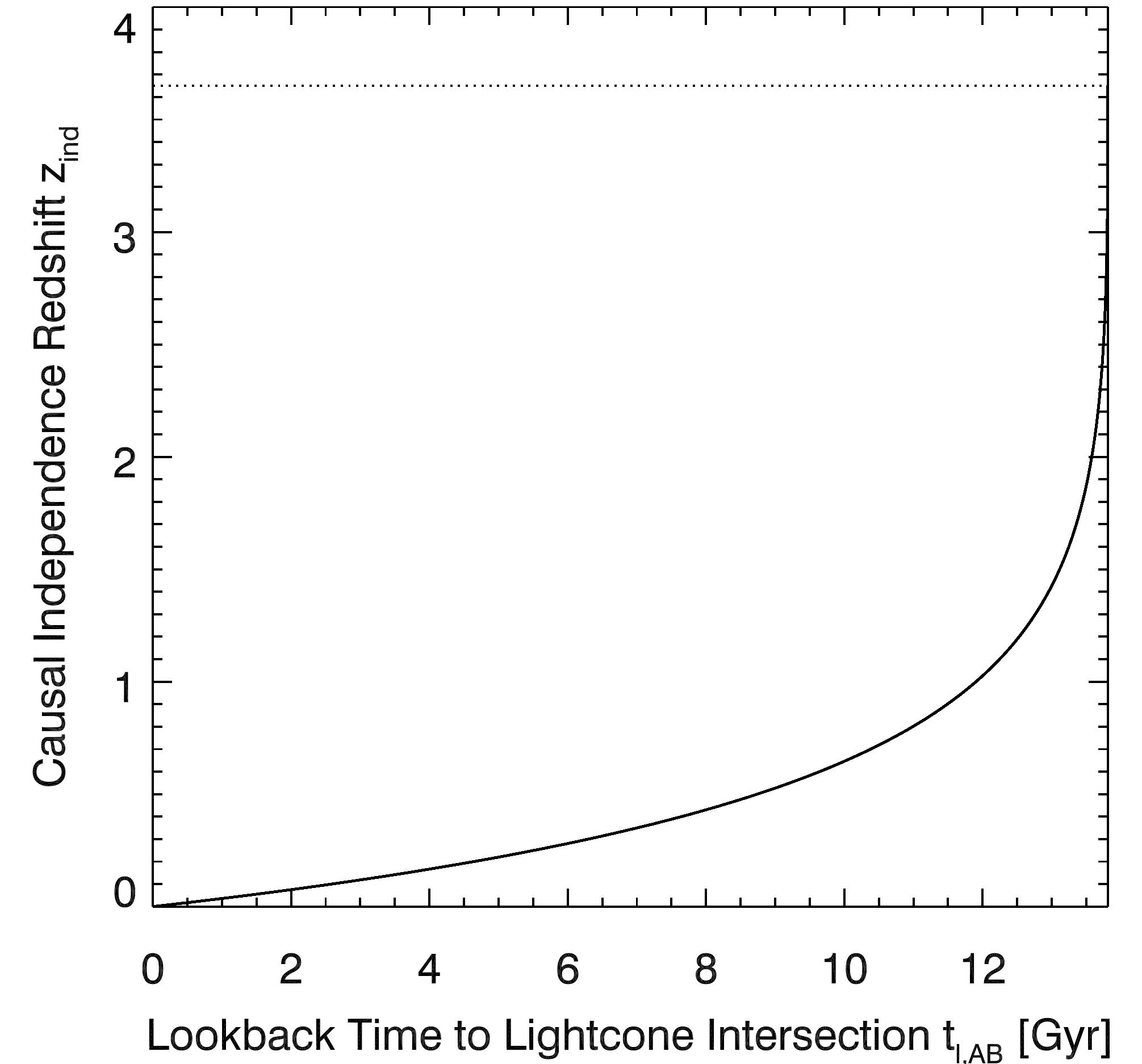}
\caption{\small \baselineskip 12pt  
({\it Left}) Redshifts $z_A$ vs. $z_B$ for the case $\alpha = 180^\circ$ corresponding to various times at which the past-directed lightcones from emission events A and B last intersected. Lightcone intersection times (in Gyr) are given in terms of conformal time since the big bang, $H_0^{-1} \tau_{AB}$, and lookback time ${t_l}_{AB}$, the cosmic time that has elapsed since the event in question. The black line toward the upper right corresponds to past-lightcone intersection at the big bang, $\tau_{AB} = 0$ as in Fig.~\ref{ABindep}. ({\it Right}) Causal-independence redshift, $z_{\rm ind}$, vs. lookback time, ${t_l}_{AB}$, for the case $z_A = z_B$ and $\alpha = 180^\circ$, which asymptotes to $z_{\rm ind} = \zindnum$ (dotted line) as the lightcone intersection approaches the time of the big bang,  ${t_l}_{AB}=\tlbb$ Gyr ago. All calculations assume parameters $\vec{\Omega}$ as in Eq. (\ref{Omegavalues}). 
}
\label{Tabindep}
\efig
%%%%%%%%%%%%%%%%%%%%%%%%%%%%

\begin{table}
\begin{center}
{\scriptsize
\begin{tabular}{|c|c|c|c|c|c|}
\hline
{\bf Event}                              & {\bf Redshift }            & {\bf Lookback Time}  & {\bf Proper Time}    &  {\bf Conformal Time}                & {\bf causal-independence redshift }\\ 
                                               &  $z$                             &  $t_{l_ {AB}}$ [Gyr]     & $t_{AB}$ [Gyr]        &  ${H_0}^{-1}\tau_{AB}$ [Gyr]   & {\bf $\tilde{z}_{\rm ind}(\tau_{AB})$}  \\
\hline
Big Bang                               & \zbb                            & $\tlbb$                          & $\tbb$                      & $\Tabb$                                      & $\zindbb$       \\
Galaxy Formed                    & $\zgal$                       & $\tlgal$                         & $\tgal$                     & $\Tabgal$                                  & $\zindgal$       \\
Earth Formed                       & $\zearth$                   & $\tlearth$                      & $\tearth$                 & $\Tabearth$                               & $\zindearth$      \\
First Eukaryotes                       & $\zeuk$                      & $\tleuk$                          & $\teuk$                      & $\Tabeuk$                                  & $\zindeuk$      \\
%First Dinosaurs                     & $\zdino$                   & $\tldino$                        & $\tdino$                   & $\Tabdino$                                & $\zinddino$      \\
%Dinosaur Extinction             & $\znodino$               & $\tlnodino$                   & $\tnodino$               & $\Tabnodino$                           & $\zindnodino$      \\
%Einstein's Birthday              & ${\zein}^{\dagger}$   & $\tlein$                         & $\tein$                     & $\Tabein$                                  & ${\zindein}^{\dagger}$     \\
\hline
\end{tabular}
}
\caption{\small \baselineskip 12pt 
Table of sample lightcone intersection times equal to times of selected past cosmic events from Fig. \ref{Tabindep}.  Redshifts $z$ in column 2 correspond to lookback, proper, and conformal times in columns 3-5.  Pushing the past-lightcone intersection event forward, $\tau_{AB} \rightarrow \tau_0$, is highly nonlinear in redshift. 
Column 6 shows the causal-independence redshift $\tilde{z}_{\rm ind}=\tilde{z}_{\rm ind}(\tau_{AB})$ for each conformal lightcone intersection time $\tau_{AB}$.
%Past cosmic events and the corresponding causal-independence redshifts, $z_{\rm ind}$, such that 
For two sources on the sky with $z_A, z_B > \tilde{z}_{\rm ind}(\tau_{AB})$ and $\alpha = 180^\circ$, the past-directed lightcones from the emission events have not intersected each other or our worldline since $\tau_{AB}$. When the past lightcones intersect at the big bang, we have the familiar $\tilde{z}_{\rm ind}(\tau_{AB}=0)=z_{\rm ind} = \zindnum$.
Computations are done for parameters $\vec{\Omega}$ from Eq. (\ref{Omegavalues}).
\label{Tab_Table}
}
\end{center}
\vspace{-0.6cm}
\end{table}

\clearpage
%----------------------------------------------------------------------------------------------------------------------------------------------------------------------------------
\section{Curved Spatial Sections}
\label{sec:curved}
%----------------------------------------------------------------------------------------------------------------------------------------------------------------------------------

We now consider how the results of Section III generalize to the cases of nonzero spatial curvature. Given the FLRW line-element in Eq. (\ref{dsgeneralchi}), radial null geodesics satisfy Eq. (\ref{nullgeochi}) for arbitrary spatial curvature $k$. For concreteness, we consider first a space of positive curvature, $k = 1$. As illustrated in Fig. \ref{SphereTriangles}, we place the Earth at point E at the north pole of the 3-sphere, with coordinates $\chi = \theta = \varphi = 0$. By construction, the coordinates $\chi$ and $\tau$ are dimensionless, while $R_0  a(\tau)$ has dimensions of length. Thus we may take the comoving spatial manifold to be a unit sphere. In that case, the coordinate $\chi_B$ (for example) gives the angle between the radial line connecting the center of the sphere (point O) to the point B on its surface, and the radial line connecting O to the point E at the north pole. Because the comoving spatial manifold has unit radius, $\chi_B$ also gives the arclength along the surface from the point B to the point E. At a given time $\tau$, the physical distance between points B and E is then given by $R_0 a(\tau) \chi_B$. See Fig. \ref{SphereTriangles}.

%%%%%%%%%%%%%%%%%%%%%%%%%%%%%%%%%%%%%%%%%%%%%%%%%%%%%%%%%%%%%%%%%
\bfig
\centering
\includegraphics[width=5in]{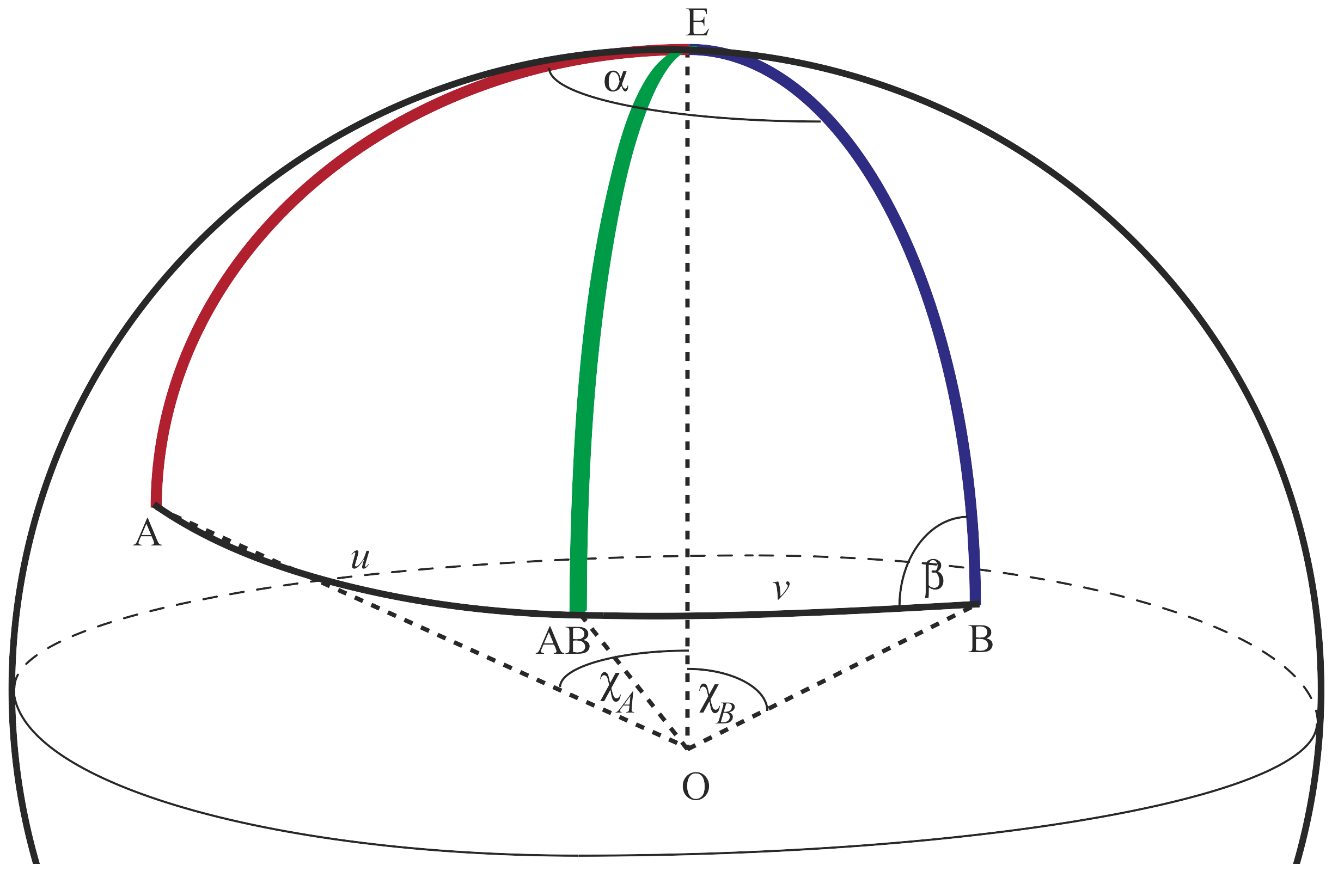}
\caption{\small \baselineskip 12pt 
The curved-space analog of Fig. \ref{Plane}b, showing emission events A and B on the unit comoving spherical manifold ($k = 1$). Earth is at the north pole (labeled point E). The center of the sphere is labeled O. The emission at event A occurs at angle $\chi_A$, which is the angle between the lines OE and OA; the emission at event B occurs at angle $\chi_B$. The past-directed lightcones from events A and B intersect at point AB, which falls along the spatial geodesic connecting points A and B. The comoving arclength between events A and AB is $u$, and the comoving arclength between events B and AB is $v$.  The angle between Earth (E) and the lightcone intersection event AB as seen from event B is $\beta$.  As usual, $\alpha$ represents the angle between emission events A and B as seen from Earth. 
%{\bf DAVE: Please label $\chi_A$. Clarify spherical nature of figure.} 
}
\label{SphereTriangles}
\efig
%%%%%%%%%%%%%%%%%%%%%%%%%%%%%%%%%%%%%%%%%%%%%%%%%%%%%%%%%%%%%%%%%

As in the spatially flat case, we take the angle (as seen from Earth) between events A and B to be $\alpha$. The past-directed lightcones from events A and B intersect at a comoving location marked AB, which falls along the spatial geodesic connecting A and B. We label the comoving arclength between points A and B as $\chi_L$; the comoving arclength from A to AB as $u$; and the comoving arclength from point AB to B as $v$, such that
%%%%%%%%%%%%
\beq
\chi_L = u + v .
\label{chiLuv}
\eeq
In our chosen coordinate system, neither A nor B is at the origin, and hence the path connecting points A and B does not appear to be a radial null geodesic. In particular, $d \theta / d\lambda \neq 0$ along the path connecting points A and B, where $\lambda$ is an affine parameter with which to parameterize the geodesic. But we may always rotate our coordinates such that point A is the new origin (at $\chi' = \theta' = \varphi' = 0$) and extend a radial null geodesic from the new origin to point ${\rm B}'$. We may then exploit the spherical symmetry of the spatial manifold to conclude that the arclength between points ${\rm A}'$ and ${\rm B}'$ will be the same as the arclength between points A and B in our original coordinate system. Thus we find that the arclength $u$ is the (comoving) radius of the past-directed lightcone between points A and AB, and from Eq. (\ref{nullgeochi}) we know that the radius of that lightcone at time $\tau_{AB}$ must equal $u = \tau_A - \tau_{AB}$. Likewise, the arclength $v = \tau_B - \tau_{AB}$. Thus Eq. (\ref{chiLuv}) is equivalent to
%%%%%%%%%
\beq
\tau_{AB} = \frac{1}{2} \left( \tau_A + \tau_B - \chi_L \right)  ,
\label{tauABk1}
\eeq
which is identical to Eq. (\ref{tauABgeneral}) for the spatially flat case. 

We next wish to relate the arclength $\chi_L$ to the inscribed angle $\alpha$.  Although Fig. \ref{SphereTriangles} is constructed explicitly for a positively curved space, we may use it to guide our application of the generalized law of cosines \cite{peebles93,peacock99} for either spherical ($k = 1$) or hyperbolic ($k = -1$) geometries. In terms of the functions $S_k (\chi)$ and $C_k (\chi)$ defined in Eqs. (\ref{Sk}) and (\ref{Ck}), the arclength $\chi_L$ between events A and B separated by an angle $\alpha$ may be written
\beqn
\label{eq:CkchiL}
C_k (\chi_L) = C_k (\chi_A) C_k(\chi_B) +  k S_k (\chi_A) S_k (\chi_B) \cos \alpha .
\eeqn
The conformal time $\tau_{AB}$ at which the past-directed lightcones intersect is thus given by Eq. (\ref{tauABk1}), with $\chi_L$ given by Eq. (\ref{eq:CkchiL}), which is equivalent to alternative expressions found in \cite{liske00,osmer81} (but see \footnote{The authors of \cite{liske00,osmer81} were primarily concerned with finding the distance $\chi_L$ between two events for arbitrary curvature separated by some angle $\alpha$ at redshifts $z_A$ and $z_B$. Those analyses were not primarily concerned with the intersection of past lightcones, although \cite{liske00} did note in passing that there existed certain angular separations $\alpha$ large enough to ensure that the past lightcones of the events at $z_A$ and $z_B$ did not intersect since the big bang (ignoring inflation).}).  

We may likewise solve for the comoving spatial coordinate, $\chi_{AB}$, at which the past-directed lightcones intersect. Using Fig. \ref{SphereTriangles}, we again label the comoving arclength from points AB to B as $v = \tau_B - \tau_{AB}$; we label the inscribed angle between arclengths $v$ and BE as $\beta$; and we use the fact that the comoving arclength from point AB to E (the green arc in Fig. \ref{SphereTriangles}) is simply $\chi_{AB}$. Then for the triangle with vertices AB, E, and B, we have, in the general curved case
\beqn
\label{eq:CkchiAB1}
C_k (\chi_{AB}) = C_k (v) C_k (\chi_B) +  k S_k (v) S_k (\chi_B) \cos \beta .
\eeqn
We may solve for the angle $\beta$ by considering the larger triangle with vertices A, B, and E, for which we may write
%%%%%%%%
\beqn
\label{eq:cosbeta2}
C_k (\chi_A) = C_k (\chi_B) C_k (\chi_L) +  k S_k (\chi_A) S_k (\chi_L) \cos \beta ,
\eeqn
where $\chi_L$ is given by Eq. (\ref{eq:CkchiL}). Using Eq. (\ref{eq:cosbeta2}) and the arclength $v = \tau_B - \tau_{AB}$, we may rearrange Eq. (\ref{eq:CkchiAB1}) to yield
\beqn
\label{eq:CkchiAB}
C_k (\chi_{AB}) = C_k (\tau_B - \tau_{AB} ) C_k (\chi_B) +  \frac{ S_k (\tau_B - \tau_{AB} ) S_k (\chi_B) } { S_k (\chi_A) S_k (\chi_L) } \left[ C_k (\chi_A) - C_k (\chi_B) C_k (\chi_L)   \right] ,
\eeqn
with $\tau_{AB}$ and $C_k (\chi_L)$ given by Eqs. (\ref{tauABk1}) and (\ref{eq:CkchiL}), respectively. 

As in the flat case ($k=0$), for the spatially curved cases ($k=\pm1$) if the past-directed lightcones from A and B intersect at time $\tau_{AB}$, given by Eq. (\ref{tauABk1}), we can 
fix $\alpha$ and $\chi_B$ to derive the condition on the critical comoving distance, $\hat{\chi}_A$,  
\beqn
\label{eq:XaTab_curved}
\hat{\chi}_A = T_{k}^{-1} \left( \frac{ C_k ( \chi_B - 2 \tau_0 + 2 \tau_{AB}  ) - C_k (\chi_B) }{ k \left[S_k (\chi_B) \cos \alpha + S_k ( \chi_B - 2 \tau_0  + 2 \tau_{AB} ) \right] } \right) ,
\eeqn
where $T_k (\chi) \equiv S_k (\chi) / C_k (\chi)$. Or we may fix $\chi_A$ and $\chi_B$ to determine the critical angle $\hat{\alpha}$ such that the past lightcones of A and B intersect at time $\tau_{AB}$,
\beqn
\label{eq:alphaTab_curved}
\hat{\alpha}  =  \cos^{-1} \left( \frac{ C_k (\tau_A + \tau_B - 2 \tau_{AB} ) - C_k (\chi_A) C_k (\chi_B) }{ k S_k (\chi_A) S_k (\chi_B) }\right) .
\eeqn
Setting $\tau_{AB} = 0$, then for $\chi_A \geq \hat{\chi}_A$ or $\alpha \geq \hat{\alpha}$ the shared causal past of the events is pushed to $\tau \leq 0$, into the inflationary epoch. We use Eq. (\ref{eq:XaTab_curved}) with $\tau_{AB} = 0$ to plot the hyperbolic curves for different angles $\alpha$ in the lefthand side of Fig. \ref{ABindepCLOSEDOPEN}, and use Eq. (\ref{chiz}) to relate $\chi$ to $z$ for the plots in the righthand side of Fig. \ref{ABindepCLOSEDOPEN}.

Eqs. (\ref{eq:XaTab_curved}) and (\ref{eq:alphaTab_curved}) are the curved-space generalizations of Eqs. (\ref{eq:XaTab_flat}) and (\ref{eq:alphaTab_flat}). It is easy to see that they reduce to the spatially flat case when $k = 0$. The limit $k \rightarrow 0$ corresponds to taking arclengths $\chi_i$ small compared to the radius of curvature. Since we are considering comoving distances on a unit comoving sphere (for $k = 1$) or on a unit hyperbolic paraboloid (for $k = -1$), the limit of interest is $\chi_i \ll 1$. Then we may use the usual power-series expansions,
%%%%%%%%%%%
\beq
\begin{split}
S_k (\chi) &= \chi + {\cal O} (\chi^3 ) , \\
C_k (\chi) &= 1 - \frac{k}{2} \chi^2 + {\cal O} (\chi^4) , \\
T_k (\chi) &= \chi + {\cal O} (\chi^3 )
\end{split}
\label{smallangle}
\eeq
to write Eqs. (\ref{eq:XaTab_curved}) and (\ref{eq:alphaTab_curved}) as
%%%%%%%%%%%%
\beq
\begin{split}
\hat{\chi}_{A} (k) &= \hat{\chi}_{A} ({\rm flat}) + {\cal O} (\chi_i^3 ) , \\
\hat{\alpha} (k) &= \hat{\alpha} ({\rm flat} ) + {\cal O} (\chi_i^4 ) 
\end{split}
\label{smallangleexpand}
\eeq
in the limit $\chi_i \ll 1$, where $\hat{\chi}_{A} ({ \rm flat )}$ and $\hat{\alpha} ({\rm flat})$ are given by Eqs. (\ref{eq:XaTab_flat}) and (\ref{eq:alphaTab_flat}), respectively.

%%%%%%%%%%%%%%%%%%%%%%%%%%%%%%%%%%%%%%%%%%%%%%%%%%%%%%%%%%%%%%%%%
\bfig
\centering
\includegraphics[width=3.2in]{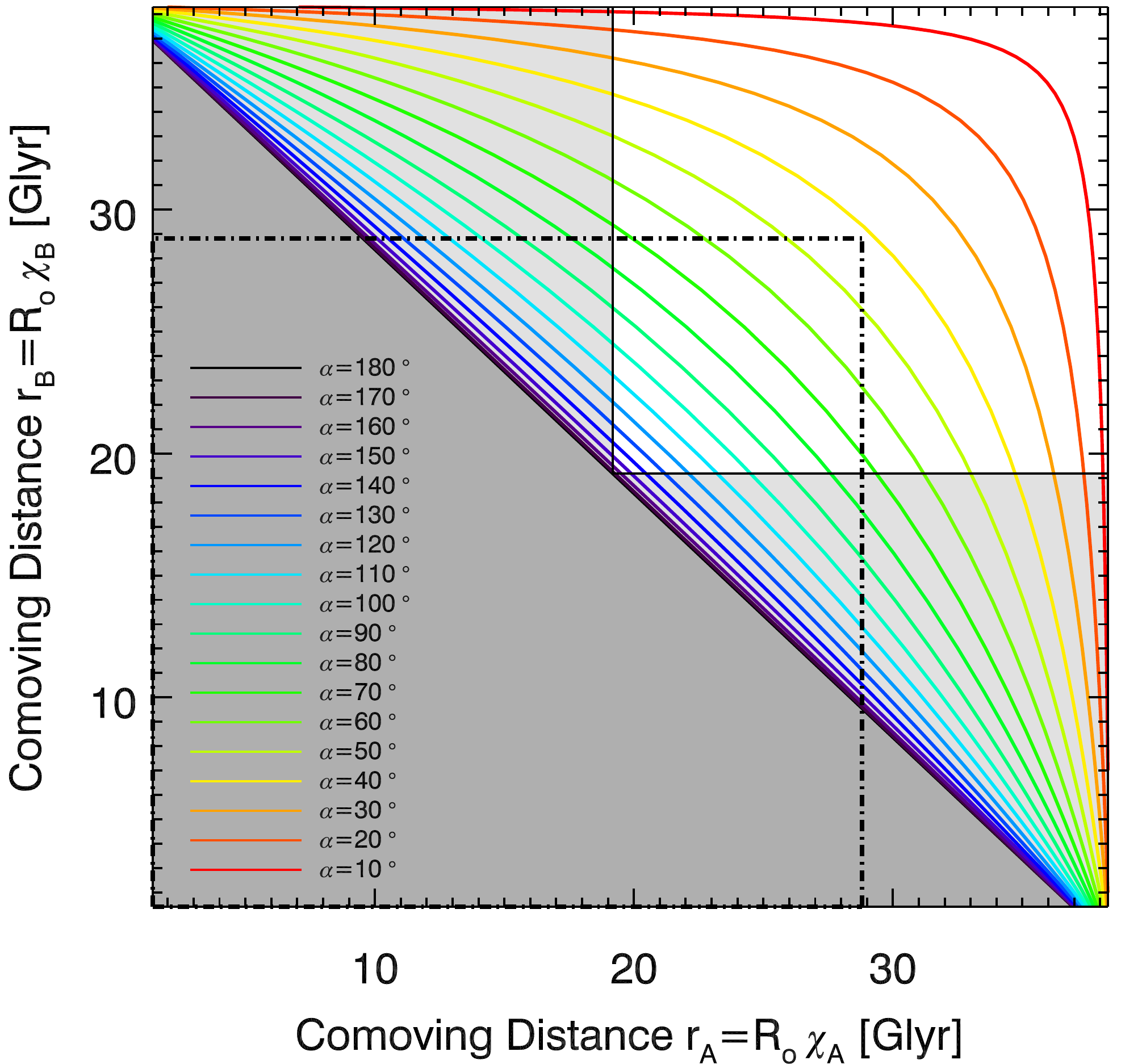}
\includegraphics[width=3.2in]{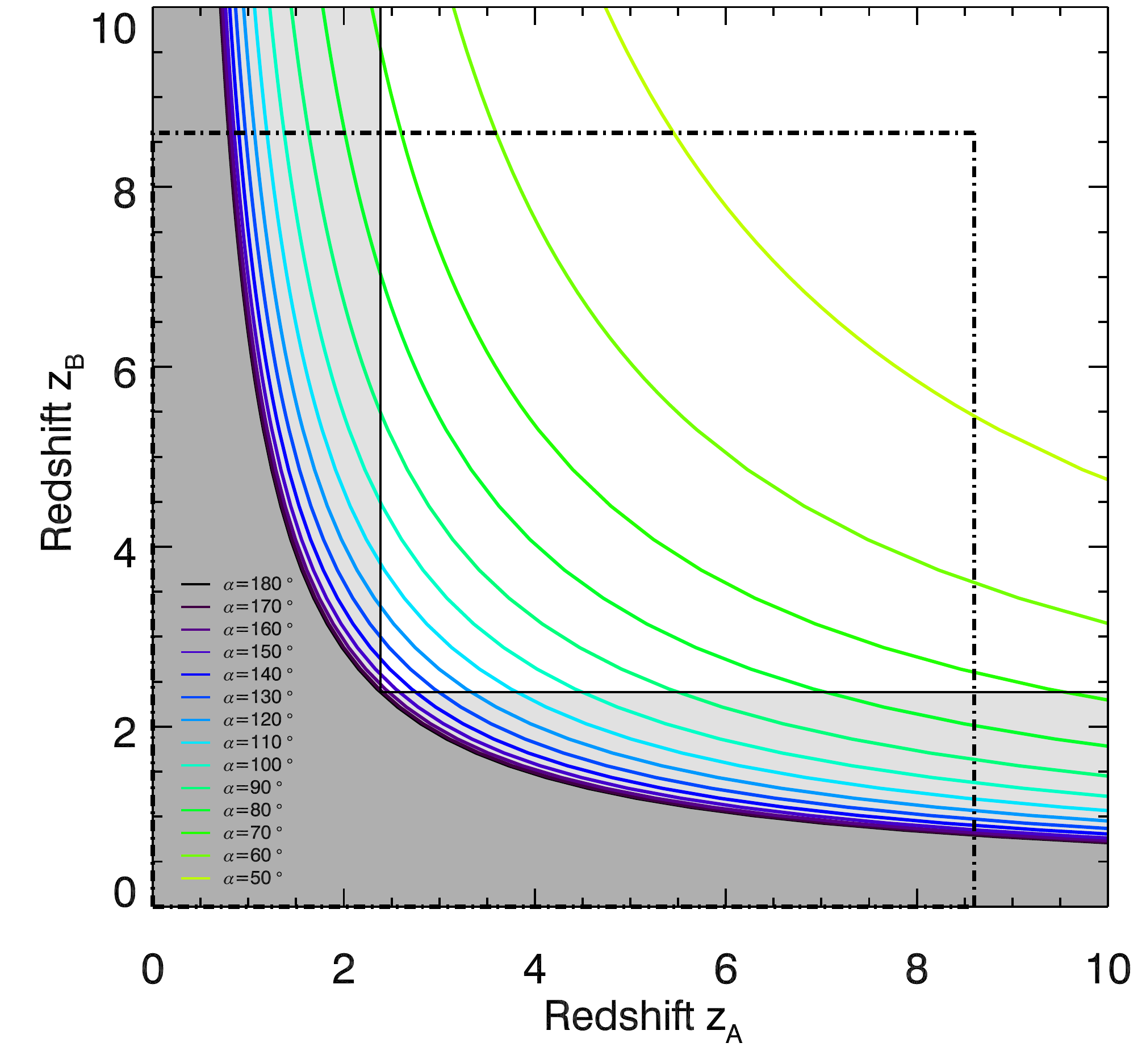}
%{lc_intersect_BB_za_zb_50_13_0_10_50p000000_180_sym_box_69p3_0p2865_0p80000001_8p6970249e-05_-0p086566982.pdf} 
\\ [5ex]
\includegraphics[width=3.2in]{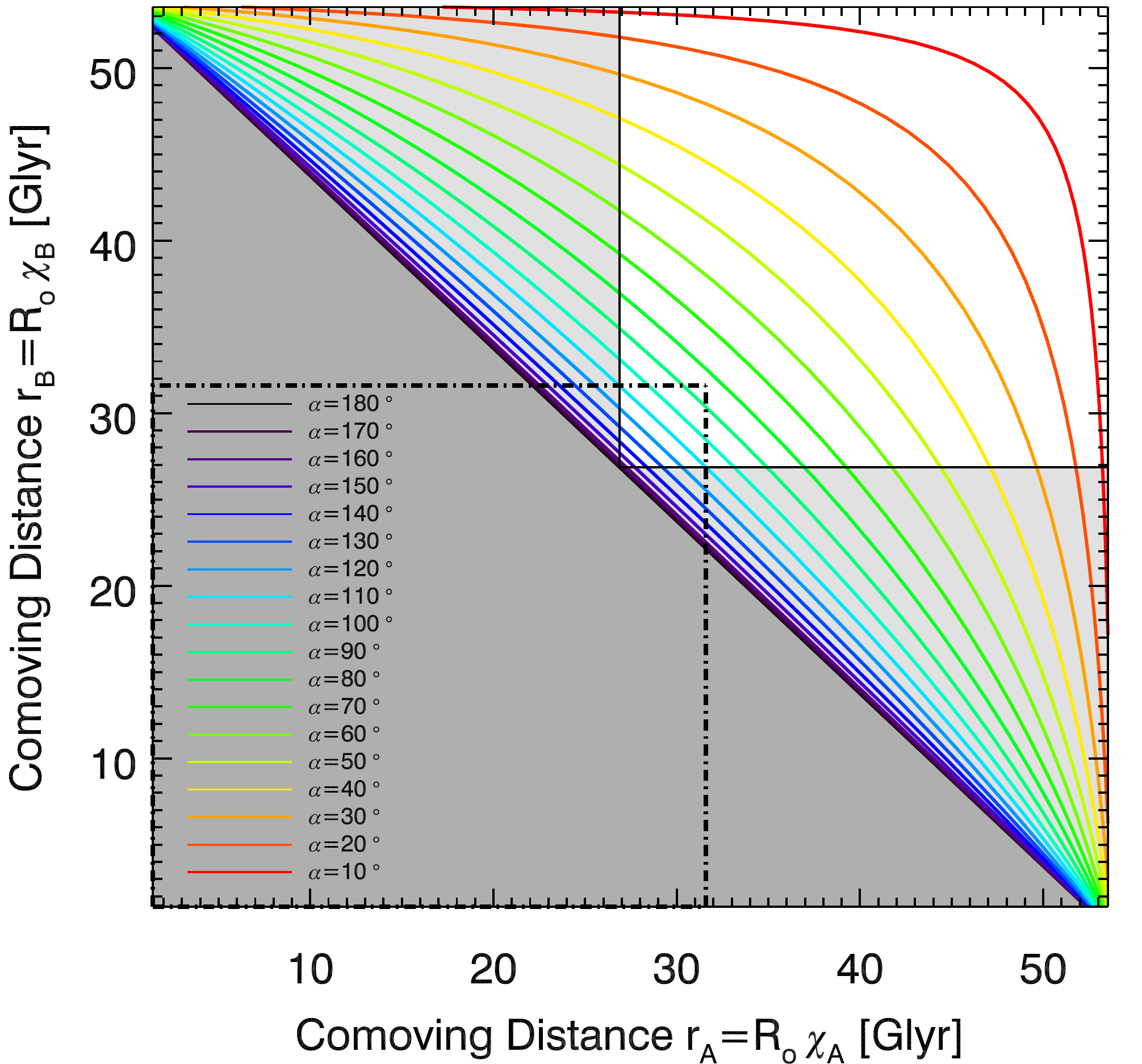}
%{lc_intersect_BB_za_zb_100_17_0p100000_10000_10p000000_180_sym_box_69p3_0p2865_0p64999998_8p6970249e-05_0p063433054.pdf}
\includegraphics[width=3.2in]{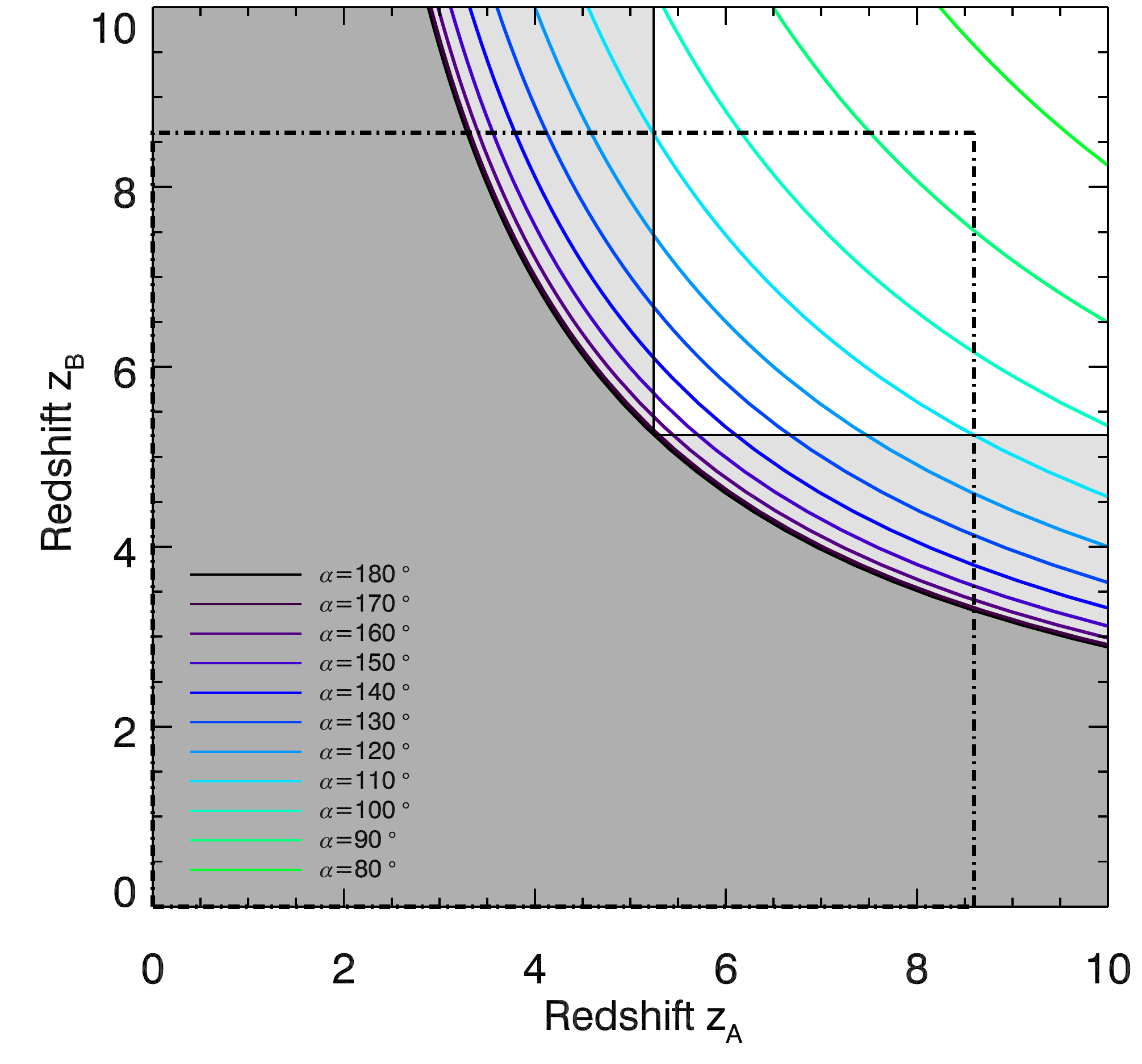}
%{lc_intersect_BB_za_zb_50_10_0_10_80p000000_180_sym_box_69p3_0p2865_0p64999998_8p6970249e-05_0p063433054.pdf}
%
\caption{\small \baselineskip 12pt
Same as Fig. \ref{ABindep} but for FLRW cosmologies with nonzero spatial curvature. We again consider parameters $\vec{\Omega} = (h, \Omega_M, \Omega_\Lambda, \Omega_R, \Omega_k, \Omega_{\rm T})$. ({\it Top Row}) A spatially closed universe ($k=1$) with $\vec{\Omega} = (\hnum, \OMnum, {\it \OLnumclosed}, \ORADnum, {\it \OKnumclosed}, {\it \OTnumclosed})$.
({\it Bottom Row}) A spatially open universe ($k = -1$) with $\vec{\Omega} = (\hnum, \OMnum, {\it \OLnumopen}, \ORADnum, {\it \OKnumopen}, {\it \OTnumopen})$. In each case, departures from the $k = 0$ case of Eq. (\ref{Omegavalues}) are indicated in italics. Compared to the $k = 0$ case, increasing $\Omega_\Lambda$ shrinks the comoving distance scale and decreases the critical redshift for a given angle, whereas decreasing $\Omega_\Lambda$ stretches the comoving distance scale and increases the critical redshift for a given angle. In all figures the dashed box represents the furthest observed object at $z_{\rm max} = \zmaxnum$, corresponding to $R_0 \chi_{\rm max} = \chimaxclosednum$ Glyr (closed), $\chimaxflatnum$ Glyr (flat), and $\chimaxopennum$ Glyr (open). The criterion that the past lightcones from events A and B do not intersect each other or our worldline for $\tau > 0$ in the $\alpha = 180^\circ$ case (white square regions in Figs~\ref{ABindep} and \ref{ABindepCLOSEDOPEN}) yields $z_A , z_B \geq \zindnumclosed$ (closed), $\zindnum$ (flat), and $\zindnumopen$ (open).
%{\bf Remake open figure for $\Omega_\Lambda=0.57$ instead of 0.65 to make it equidistant from 0.8 and $\OLnum$.}
}
\label{ABindepCLOSEDOPEN}
\efig
%%%%%%%%%%%%%%%%%%%%%%%%%%%%%%%%%%%%%%%%%%%%%%%%%%%%%%%%%%%%%%%%%

Comparing Figs. \ref{ABindep} and \ref{ABindepCLOSEDOPEN}, one finds that FLRW universes with the same values of $\Omega_M$ and $\Omega_R$ as ours but with different values of $\Omega_\Lambda$ yield different values of the critical angle $\hat{\alpha}$ at which objects with redshifts $z_A$ and $z_B$ satisfy $\tau_{AB} \leq 0$. First note that $\Omega_{\Lambda,f} = \OLnum$ is the value of $\Omega_\Lambda$ in Eq. (\ref{Omegavalues}) corresponding to our Universe. For a closed universe ($\Omega_{\Lambda} > \Omega_{\Lambda,f}$) the range of critical angles $\hat{\alpha}$ for which one may find objects with redshifts $z_A$ and $z_B$ that satisfy the condition $\tau_{AB} \leq 0$ is broader than in the spatially flat case, whereas in an open universe ($\Omega_\Lambda < \Omega_{\Lambda,f}$) the range of critical angles $\hat{\alpha}$ is narrower than in the spatially flat case. These results are exactly as one would expect given the effect on the inscribed angle $\alpha$ at the point E as one shifts from a Euclidean triangle ABE to a spherical triangle or a hyperbolic triangle.

\clearpage
%----------------------------------------------------------------------------------------------------------------------------------------------------------------------------------
\section{Future Lightcone Intersections}
\label{sec:disc}
%----------------------------------------------------------------------------------------------------------------------------------------------------------------------------------

To extend our analysis of shared causal domains to the future of events A and B we define $\tau_\infty$, the total conformal lifetime of the Universe,
\beqn
\label{eq:tau_inf}
\tau_\infty \equiv \tau (t = \infty) = \int_{0}^{\infty} \frac{da}{a^2 E(a)} .
\eeqn
As usual, $\tau_\infty$ is dimensionless while $R_0 \tau_\infty / c = H_0^{-1} \tau_\infty$ is measured in Gyr. We restrict attention to cosmologies like our own ($\Lambda$CDM with $k = 0$ and $\Omega_\Lambda > 0$) that undergo late-time cosmic acceleration and expand forever; that ensures that the total conformal lifetime of the universe is {\it finite}, $\tau_\infty < \infty$. In particular, for $\vec{\Omega}$ as in Eq. (\ref{Omegavalues}), we find $H_0^{-1} \tau_\infty  = \tauinfnum$ Gyr.  See Fig.~\ref{Conformal} and Fig.~\ref{fig:Hub}.

FLRW cosmologies with a finite conformal lifetime necessarily have cosmic event horizons \cite{peacock99}. Objects we observe today that are beyond the cosmic event horizon have already emitted the last photons that will ever reach us (at $t=\infty$), and it is impossible for us to send a signal today that will ever reach those objects in the future history of our Universe \cite{starobinsky00,starkman99,loeb02,davis04}. The condition $\tau_\infty < \infty$ holds for FLRW cosmologies with nonzero spatial curvature ($k \neq 0$) as long as $\Omega_\Lambda > 0$ is large enough that dark energy domination sets in before matter, curvature, or radiation domination causes the universe to re-collapse
\footnote{We do not consider certain FLRW universes with different types of event horizons. Although universes that eventually recollapse in a ``big crunch'' (for positive or negative $\Omega_\Lambda$) also have a finite conformal lifetime and thus an event horizon, they also have a finite proper lifetime and do not expand forever.  Similarly, we do not consider bouncing or loitering cosmologies with $\Omega_\Lambda > 0$ that have a maximum observable redshift and did not have a ``big bang'' in the past.  These cosmologies do expand forever and can have event horizons but were contracting in the past followed by expansion from a repulsive bounce due to a non-zero value of dark energy}.

%%%%%%%%%%%%%%
\bfig 
\centering
\includegraphics[width=6.5in]{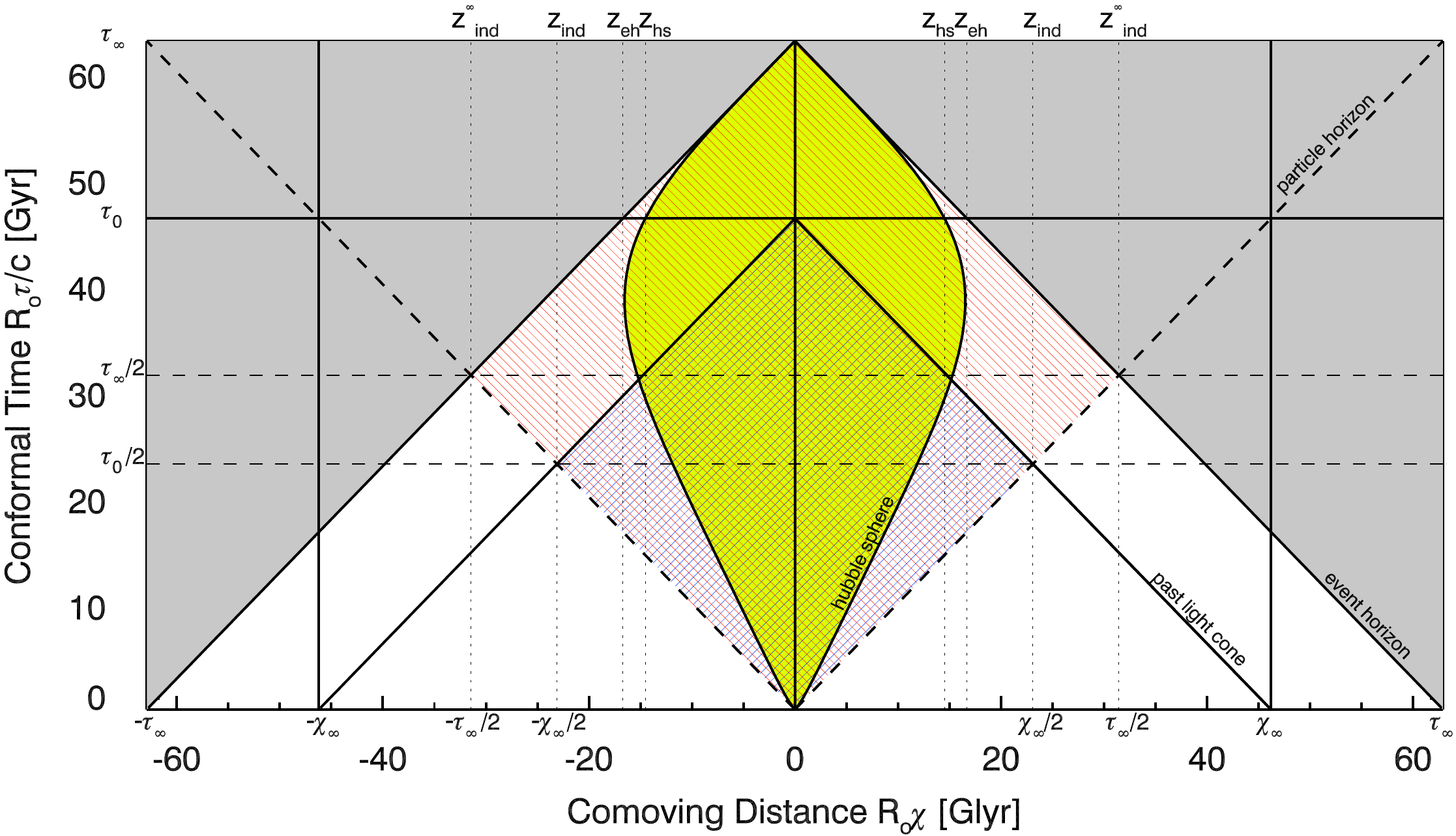}
\caption{\small \baselineskip 12pt 
Conformal diagram as in Fig.~\ref{Conformal} showing the causal independence region bounded by the particle horizon and the past-directed lightcone from the present time, $\tau_0$ (purple cross-hatching); the causal diamond bounded by the particle horizon and the cosmic event horizon (red stripes tilted at -45 degrees), which includes the causal independence region; and the Hubble sphere (equal to the apparent horizon for $\Omega_k = 0$; see Appendix B), which is the spacetime region beyond which all objects are receding faster than light (yellow). Relevant redshifts include the current value of the redshift of the Hubble sphere, $z_{\rm hs} = \zhsnum$; the current redshift of the event horizon, $z_{\rm eh} = \zehnum$; the current value of the causal-independence redshift, $z_{\rm ind} = \zindnum$; and the current value of the redshift that bounds the causal diamond, $z_{\rm ind}^\infty = \zindinfnum$, which is the limiting value of the causal-independence redshift as the proper age of the universe approaches infinity. 
\label{fig:Hub}
}
\efig
%%%%%%%%%%%%%%%%%%%%%%%%%%%%%%%%%%%%%%%%%%%%%%%%%%%%%%%%%%%%%%%%%

The event horizon is a particular past-directed lightcone, and hence the surface is a null geodesic. Thus we may use Eq. (\ref{chiz}), suitably modifying the limits of integration. At a particular time, $a_* = a (t_*)$, the comoving distance from our worldline at $\chi = 0$ to the event horizon is given by
%%%%%%%%%
\beq
\chi_{\rm eh} (t_*) = \int_{a_*}^\infty \frac{da}{a^2 E (a) } .
\label{eq:chieh1}
\eeq
We may also trace back along the past lightcone from our present location (at $\tau_0$ rather than $\tau_\infty$) to the equivalent comoving distance. We set $a (t_*) = a(t_0) = 1$ and compute
%%%%%%
\beq
\chi (t_0) = \int_{a_{\rm eh} }^1 \frac{da}{a^2 E(a) } .
\label{eq:chieh2}
\eeq
Equating Eqs. (\ref{eq:chieh1}) and (\ref{eq:chieh2}) and using $z_{\rm eh} = a_{\rm eh}^{-1} - 1$, we find $z_{\rm eh} (t_0) = \zehnum$ for our cosmology with $\vec{\Omega}$ as in Eq. (\ref{Omegavalues}). Note that since $z_{\rm eh} < z_{\rm ind} = \zindnum$, objects with $z \geq z_{\rm ind}$ are beyond the cosmic event horizon: though we have received light from them at $\tau_0$, no return signal from us will ever reach them before $\tau_\infty$, nor (symmetrically) can light emitted from them now (at $\tau_0$) ever reach us before the end of time. See Fig.~\ref{Conformal} and Fig.~\ref{fig:Hub}.

Another quantity of interest is the value of the redshift today of an emission event whose light we will receive at $\tau_\infty$ but whose past lightcone has no overlap with our worldline since $\tau = 0$. Such will be the case for any object with redshift $z > z_{\rm ind}^\infty$. As can be seen from Fig. \ref{fig:Hub}, $z_{\rm ind}^\infty$ corresponds to the comoving location where the cosmic event horizon intersects the future lightcone from the origin, namely at the spacetime point $(\chi, \tau ) = (\tau_\infty / 2 , \tau_\infty / 2)$. We may therefore evaluate $z_{\rm ind}^\infty$ either by computing the comoving distance from the origin to the event horizon at $\tau_\infty / 2$, or by computing the comoving distance of the forward lightcone from the origin at $\tau_\infty / 2$. In the first case we have
%%%%%%%
\beq
\chi_{\rm eh} \left(\frac{ \tau_\infty }{2} \right) = \left( \tau_\infty - \frac{\tau_\infty}{2} \right) =  \int_{a_{\rm ind}^\infty}^{\infty} \frac{da}{a^2 E(a)} ,
\label{eq:chiehtau2}
\eeq
and in the second case we have
%%%%%%%
\beq
\chi_{\rm flc} \left( \frac{ \tau_\infty}{ 2} \right) = \left(  \frac{\tau_\infty}{ 2} - 0 \right) = \int_{0}^{a_{\rm ind}^\infty} \frac{da}{a^2 E (a) } .
\label{eq:chiehflc}
\eeq
Numerically inverting either Eq. (\ref{eq:chiehtau2}) or (\ref{eq:chiehflc}) and using $z_{\rm ind}^\infty = (a_{\rm ind}^\infty)^{-1} -1$, we find $z_{\rm ind}^\infty = \zindinfnum > z_{\rm ind}$ for our cosmology with $\vec{\Omega}$ as in Eq. (\ref{Omegavalues}). We emphasize that both $z_{\rm ind}^\infty$ and $z_{\rm ind}$ are evaluated at the time $\tau_0$: among the objects whose redshift we might measure today, those with $z > z_{\rm ind}^\infty$ will (later) release light that will reach our worldline at $\tau_\infty$ and whose past lightcones from that later emission event will have had no overlap with our worldline since $\tau = 0$. 
 
Events have no shared causal future if their future lightcones will never intersect each other's worldlines before $\tau_{\infty}$. Thus we may ask whether the forward lightcone from emission event A intersects with the worldline of event B at some time $\tau_0 < \tau \leq \tau_\infty$, or vice versa. This question can be answered by visual inspection of Fig.~\ref{Conformal} for the special case for our universe when $\alpha=180^{\circ}$ with fixed redshifts $z_A=1$, $z_B=3$.  In Fig.~\ref{Conformal}, the future lightcones from events A and B are shown as thin dashed lines, and the worldines of A and B are shown as thin dotted lines at the fixed comoving locations $\chi_A$ and $\chi_B$, respectively.  From Fig.~\ref{Conformal}, it is easy to see that the future lightcone from event B crosses event A's worldline before $\tau_{\infty}$ while the future lightcone from event A does not cross event B's worldline before $\tau_{\infty}$.  Thus, in this situation, event B can send a signal to the comoving location of event A before the end of time, while event A can never signal event B's worldline even in the infinite future.  Similarly, we can consider the future lightcone from Earth today in Fig. \ref{Conformal}, and note that, while we can signal the comoving location of event A before time ends, we will never be able to send a signal that will reach the comoving location of event B.  Of course, as shown in Fig.~\ref{Conformal}, events A and B have already signaled Earth by virtue of our observing their emission events along our past lightcone at $(\chi,\tau)=(0,\tau_0)$, and the future lightcone from Earth today necessarily overlaps with the future lightcones of events A and B for $\tau > \tau_0$. 

For general cases at different angles and redshifts, without loss of generality we retain the condition that emission event A occurred later than B, $\tau_A \geq \tau_B$. We introduce the notation that $\tilde{\tau}_{ij}$ is the conformal time when the future lightcone from event $i$ intersects the worldline of event $j$, for $\tilde{\tau}_{ij} > \tau_0$. Using Fig. \ref{Conformal} and reasoning as in Sections \ref{sec:flat} and \ref{sec:curved}, we find
\beqn
\label{eq:Tab_future}
\begin{split}
\tilde{\tau}_{AB} &= \chi_L + \tau_A , \\
\tilde{\tau}_{BA} &= \chi_L + \tau_B ,
\end{split} 
\eeqn
where $\chi_L$ is the comoving distance between events A and B given by Eqs. (\ref{rL}) and (\ref{eq:CkchiL}) for the spatially flat and curved cases, respectively. Since all angular and curvature dependence is implicit in the $\chi_L$ term, Eq. (\ref{eq:Tab_future}) holds for arbitrary angular separations $0 \leq \alpha \leq 180^{\circ}$ and curvatures ($k=0,\pm 1$). In general $\tilde{\tau}_{AB} \neq \tilde{\tau}_{BA}$; the two are equal only if $\tau_A = \tau_B$. Given our assumption that $\tau_A \geq \tau_B$ it follows that $\tilde{\tau}_{AB} \geq \tilde{\tau}_{BA}$. 

Three scenarios are possible. $( {\it a} )$ Events A and B will each be able to send a light signal to the other, $\tilde{\tau}_{BA} \leq \tilde{\tau}_{AB} < \tau_\infty$, which implies $\chi_L < \tau_\infty - \tau_A \leq \tau_\infty - \tau_B $. $( {\it b} )$ B will be able to send a signal to A but not vice versa, $\tilde{\tau}_{BA} < \tau_\infty < \tilde{\tau}_{AB}$, which implies $\tau_\infty - \tau_A < \chi_L < \tau_\infty - \tau_B$. $( {\it c })$ A and B will forever remain out of causal contact with each other, $\tilde{\tau}_{AB} \geq \tilde{\tau}_{BA} \geq \tau_\infty$, which implies $\tau_\infty - \tau_A \leq \tau_\infty - \tau_B < \chi_L$.

Fixing $\chi_B$ and $\alpha$, we may find the comoving distance $\tilde{\chi}_A$ such that the future lightcone from A will intersect the worldline of B at time $\tilde{\tau}_{AB}$. For a spatially flat universe ($k = 0$), we find
%%%%%%%
\beq
\tilde{\chi}_A = \frac{ \chi_B^2 - ( \tilde{\tau}_{AB} - \tau_0 )^2 }{2 \left( \tilde{\tau}_{AB} - \tau_0 + \chi_B \cos \alpha \right) } .
\label{eq:XaTabFflat}
\eeq
Or we may fix $\chi_A$ and $\chi_B$ and find the critical angle, $\tilde{\alpha}_{AB}$, such that the future lightcone from A intersects the worldline of B at time $\tilde{\tau}_{AB}$,
%%%%%%%%%
\beq
\tilde{\alpha}_{AB} = \cos^{-1} \left( \frac{ \chi_A^2 + \chi_B^2 - ( \tilde{\tau}_{AB} - \tau_A )^2 }{2 \chi_A \chi_B } \right) .
\label{eq:alphaTabFflat}
\eeq
As in Section \ref{sec:curved}, we may generalize these results to the case of spatially curved geometries ($k = \pm 1$), to find
%%%%%%%
\beq
\tilde{\chi}_A = T_k^{-1} \left( \frac{ C_k ( \tilde{\tau}_{AB} - \tau_0 ) - C_k (\chi_B ) }{ k \left[ S_k (\chi_B) \cos \alpha + S_k (\tilde{\tau}_{AB} - \tau_0 ) \right] } \right)
\label{eq:XaTabFcurved}
\eeq
and
%%%%%%%%%%
\beq
\tilde{\alpha}_{AB} = \cos^{-1} \left( \frac{ C_k (\tilde{\tau}_{AB} + \tau_A ) - C_k (\chi_A ) C_k (\chi_B ) }{ k S_k (\chi_A ) S_k (\chi_B ) } \right) .
\label{eq:alphaTabFcurved}
\eeq
For Eqs. (\ref{eq:XaTabFflat})--(\ref{eq:alphaTabFcurved}), the comparable expressions ($\tilde{\chi}_B$ and $\tilde{\alpha}_{BA}$) for the case in which the future lightcone from B intersects the worldline of A at time $\tilde{\tau}_{BA}$ follow upon substituting $\chi_B \longleftrightarrow \chi_A$, $\tau_B \longleftrightarrow \tau_A$, and $\tilde{\tau}_{AB} \rightarrow \tilde{\tau}_{BA}$.

With these expressions in hand, we may draw general conclusions about whether events A and B share a causal past and/or a causal future. From Eq. (\ref{tauABgeneral}), the condition for no shared causal past since the big bang, $\tau_{AB} \leq 0$, is equivalent to
%%%%%%
\beq
\tau_A + \tau_B \leq \chi_L ,
\label{nopast}
\eeq
while from Eq. (\ref{eq:Tab_future}), the condition that A and B share no causal future, $\tilde{\tau}_{BA} \geq \tau_\infty$, is equivalent to
%%%%%%%%
\beq
\tau_\infty - \tau_B \leq \chi_L .
\label{nofuture}
\eeq
Each of these conditions holds for arbitrary spatial curvature and angular separation, provided one uses the appropriate expression for $\chi_L$, Eq. (\ref{rL}) or (\ref{eq:CkchiL}). Thus the criterion that events A and B share neither a causal past nor a causal future between the big bang and the end of time is simply
%%%%%%%%%
\beq
\tau_A + \tau_B < \chi_L \>\> {\it and} \>\> \tau_\infty - \tau_B < \chi_L .
\label{nopastnofuture}
\eeq
If instead
%%%%%%%%%%%
\beq
\tau_A + \tau_B < \chi_L < \tau_\infty - \tau_B ,
\label{nopastyesfuture}
\eeq
then events A and B share no causal past but B will be able to signal A in the future. And if
%%%%%%%%
\beq
\tau_\infty - \tau_B < \chi_L < \tau_A + \tau_B ,
\label{yespastnofuture}
\eeq
then events A and B share no causal future though their past lightcones did overlap after the big bang. 

If we further impose the restriction that events A and B share no past causal with each other {\it or} with our worldline, hence $z_A, z_B \geq z_{\rm ind} > z_{\rm eh}$, then by necessity events A and B will share no causal future, nor will we be able to send a signal to either event's worldline before the end of time. The reason is simple: too little (conformal) time remains between $\tau_0$ and $\tau_\infty$. Our observable universe has entered late middle-age: as measured in conformal time, the present time, $H_0^{-1} \tau_0 = \tauonum$ Gyr, is considerably closer to $H_0^{-1} \tau_\infty  = \tauinfnum$ Gyr than to the big bang at $H_0^{-1} \tau = 0$. That conclusion could change if the dark energy that is causing the present acceleration of our observable universe had an equation of state different from $w = -1$. In that case, $\Omega_\Lambda$ would vary with time and thereby alter the future expansion history of our universe.

%----------------------------------------------------------------------------------------------------------------------------------------------------------------------------------
\section{Conclusions}
\label{sec:conc}
%----------------------------------------------------------------------------------------------------------------------------------------------------------------------------------

We have derived conditions for whether two cosmic events can have a shared causal past or a shared causal future, based on the present best-fit parameters of our $\Lambda$CDM cosmology. We have further derived criteria for whether either cosmic event could have been in causal contact with our own worldline since the big bang (which we take to be the end of early-universe inflation \cite{guth05,bassett06}); and whether signals sent from either A or B could ever reach the worldline of the other during the finite conformal lifetime of our universe. We have derived these criteria for arbitrary redshifts, $z_A$ and $z_B$, as well as for arbitrary angle $\alpha$ between those events as seen from Earth. We have also derived comparable criteria for the shared past and future causal domains for spatially curved FLRW universes with $k = \pm 1$.

For the best-fit parameters of our $\Lambda$CDM cosmology, we find that if emission events A and B appear on opposite sides of the sky ($\alpha = 180^\circ$), then they will have been causally independent of each other and our worldline since the big bang if $z_A, z_B > z_{\rm ind} = \zindnum$. More complicated relationships between $z_A$ and $z_B$ must be obeyed to maintain past causal independence in the case of $\alpha < 180^\circ$, as illustrated in Fig. \ref{ABindep}b. Observational astronomers have catalogued tens of thousands of objects with redshifts $z > \zindnum$ (see, e.g., \cite{schneider10,paris12,flesch12}), and we have presented sample pairs of quasars that satisfy all, some, or none of the relevant criteria for vanishing past causal overlap with each other and with our worldline since the time of the big bang (Fig. \ref{ABempirical} and Table \ref{QTable}). Likewise, because of  non-vanishing dark energy, our observable universe has a finite conformal lifetime, $\tau_\infty$, and hence a cosmic event horizon. Our present time $\tau_0$ is closer to $\tau_\infty$ than to $\tau = 0$. Events at a current redshift of $z > \zehnum$ are beyond the cosmic event horizon, and no signal sent from us today will ever reach their worldline. Symmetrically, objects currently at $z=1.87$ are just now sending the last photons that will ever reach us in the infinite future.

Throughout our analysis we have defined $\tau = 0$ to be the time when early-universe inflation ended (if inflation indeed occurred). If there were a phase of early-universe inflation for $\tau < 0$ that persisted for at least $\sim 65$ efolds, as required to solve the flatness and horizon problems \cite{guth05,bassett06}, then {\it all} events within our past lightcone would have past lightcones of their own that intersect during inflation (see Appendix A). Based on our current understanding of inflation, however, the energy that drove inflation must have been transformed into the matter and energy of ordinary particles at the end of inflation in a process called ``reheating" \cite{guth05,bassett06,kofman08,allahverdi10}. In many models, reheating (and especially the phase of  explosive ``preheating") is a chaotic process for which --- in the absence of new physics --- it is difficult to imagine how meaningful correlations between specific cosmic events A and B, whose past lightcones have not intersected since the end of reheating, could survive to be observable today. We therefore assume that emission events A and B whose only shared causal past occurs during the inflationary epoch have been {\it effectively} causally disconnected since $\tau > 0$. 

However, if correlations between inflationary era events somehow did survive to be observable today via later emission events at their comoving locations, we could quantify the shared causal history of such events with the formalism presented here, extended to include inflation.  Certainly, in the standard cosmological view, inflationary era correlations are observable today as statistical patterns in the CMB, imprinted by inflationary era quantum fluctuations which seed present day large scale cosmic structure \cite{guth05,ade13}. While the observable spatial densities of galaxies, clusters, and thus quasars are thought to reflect correlations set up during inflation, it remains an open question whether inflationary era events at specific comoving locations --- {\it where quasar host galaxies later formed} --- could yield an observable correlation signal between pairs of eventual quasar {\it emission events} at those same comoving locations billions of years after the inflationary density perturbations were imprinted.

In closing, we note that all of our conclusions are based on the assumption that the expansion history of our observable universe, at least since the end of inflation, may be accurately described by canonical general relativity and a simply-connected, non-compact FLRW metric. These assumptions are consistent with the latest empirical search for non-trivial topology, which found no observable signals of compact topology for fundamental domains up to the size of the surface of last scattering \cite{PlanckTopology13}. 

Future work will apply our results to astrophysical data by searching the Sloan Digital Sky Survey database \cite{schneider10,paris12} and other quasar datasets comprising more than one million observed quasars \cite{flesch12} to identify the subset of pairs whose past lightcones have not intersected each other or our worldline since the big bang at the end of inflation. We also note that though the results in this paper were derived for pairs of cosmic events, they may be extended readily  to larger sets of emission events by requiring that each pairwise combination satisfies the criteria derived here. Applying the formalism developed here, using best-fit $\Lambda$CDM parameters, to huge astrophysical datasets will enable physicists to design realistic experiments of fundamental properties that depend upon specific causal relationships.  This is of particular importance for quantum mechanical experiments that crucially depend on whether certain physical systems are prepared independently.  Many such experiments implicitly assume preparation independence of subsystems even though such systems demonstrably have a fairly recent shared causal past, extending back only a few milliseconds for Earth-bound systems.  This work will allow experimenters to identify cosmological physical systems with emission events that have been causally independent for billions of years, including emission event pairs that are as independent as the expansion history of the universe will allow on causal grounds alone, modulo any shared causal dependence set up during inflation.  Future experiments which observe such causally disjoint astronomical sources may allow us to leverage cosmology to test fundamental physics including different aspects of quantum mechanics, specific models of inflation, and perhaps even features of a future theory of quantum gravity.

%----------------------------------------------------------------------------------------------------------------------------------------------------------------------------------
%\section{Appendix}
%\label{sec:appendix}
%----------------------------------------------------------------------------------------------------------------------------------------------------------------------------------

\section*{Appendix A. Inflation and the Horizon Problem}
\label{sec:horizon}
%----------------------------------------------------------------------------------------------------------------------------------------------------------------------------------

%{\bf redo calculations with Planck cosmo parameters}

Using Eq. (\ref{eq:alphaTab_flat}) and $\vec{\Omega}$ from Eq. (\ref{Omegavalues}), we may solve for the critical angular separation $\hat{\alpha}_{\rm CMB}$ at the redshift of CMB formation ($z_{\rm CMB} =\zcmbnum$ \cite{ade13}), when matter and radiation decoupled. For $z_A = z_B = z_{\rm CMB}$, and therefore $\chi_A = \chi_B = \chi_{\rm CMB}$ and $\tau_A = \tau_B = \tau_{\rm CMB}$, we find from Eq. (\ref{eq:alphaTab_flat})
\beqn
\label{eq:alphaBB_CMB}
\hat{\alpha}_{\rm CMB}  =   \cos^{-1} \left[ 1 - 2 \left(\frac{\tau_{\rm CMB} }{\chi_{\rm CMB} }\right)^2 \right] = 2 \sin^{-1} \left( \frac{\tau_{\rm CMB} }{\chi_{\rm CMB} } \right) . 
\eeqn
Using $z_{\rm CMB} = \zcmbnum$ and evaluating $\chi_{\rm CMB}$ and $\tau_{\rm CMB}$ using Eqs. (\ref{chiz}) and (\ref{eq:tau}), then Eq. (\ref{eq:alphaBB_CMB}) yields $\hat{\alpha}_{\rm CMB} = \alphacmbnum^{\circ}$. Without inflation, CMB regions on the sky that we observe today with an angular separation $\hat{\alpha}_{\rm CMB} > \alphacmbnum^{\circ}$ could not have been in causal contact at the time when the CMB was emitted. Our formalism considers the angle $\alpha$ between events A and B as seen from Earth. At a given time, $\tau$, the particle horizon subtends an angle $\theta = \alpha  / 2$ as seen from Earth, and hence our result is equivalent to the one commonly reported in the literature, $\hat{\theta}_{\rm CMB} = 1.16^\circ$ \cite{bassett06}.

If early-universe inflation did occur, on the other hand, then the past lightcones for such regions could overlap at times $\tau < 0$. We may calculate the minimum duration of inflation required to solve the horizon problem. The conformal time that has elapsed between the release of the CMB and today is $\tau_0 - \tau_{\rm CMB}$. In order to guarantee that all regions of the CMB that we observe today could have been in causal contact at earlier times, we require 
%%%%%%%%%%%
\beq
\Delta \tau_{\rm infl} + \tau_{\rm CMB} \geq \tau_0 - \tau_{\rm CMB} ,
\label{Deltatauinfl1}
\eeq
where $\Delta \tau_{\rm infl}$ is the duration of inflation in (dimensionless) conformal time. The condition in Eq. (\ref{Deltatauinfl1}) ensures that the forward lightcone from $\chi = 0$ at the beginning of inflation, $\tau_i$, encompasses the entire region of the $\tau_{\rm CMB}$ hypersurface observable from our worldline today. In the notation of Sections \ref{sec:flat}-\ref{sec:curved}, this is equivalent to setting the time at which the past lightcones from the distant CMB emission events intersect, $\tau_{AB}$, equal to the start of inflation, $\tau (t_i)$, or $\tau_{AB} = \tau (t_i) < 0$. See Fig. \ref{fig:Inflation}. 

%%%%%%%%%%%%%%
\bfig 
\centering
\includegraphics[width=6.5in]{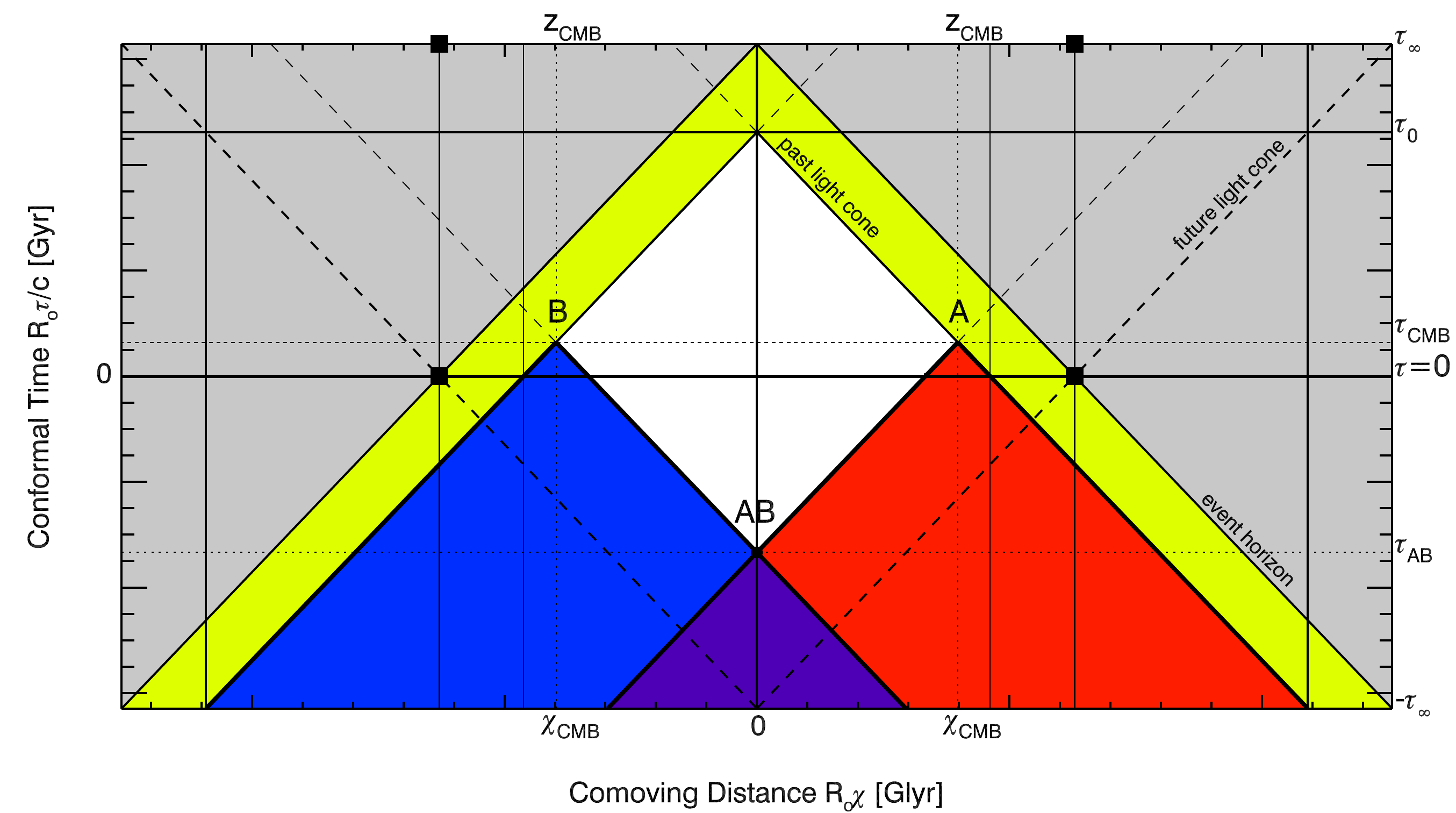}
\caption{\small \baselineskip 12pt 
Conformal diagram illustrating how inflation solves the horizon problem.  Two CMB emission events A and B are shown on opposite sides of the sky at $z_A = z_B = z_{\rm CMB}$. The region bounded by the four filled black squares is the conformal diagram without inflation, akin to Fig. \ref{Conformal}, showing that the past lightcones from events A and B (red and blue triangles, respectively) do not intersect since the big bang at $\tau=0$ (thick black horizontal line). With inflation, the diagram extends to negative conformal times, $\tau < 0$. If inflation persists for at least $\Delta \tau_{\rm infl} = \vert \tau_{\rm AB} \vert \geq \tau_0 - 2 \tau_{\rm CMB}$, then the forward lightcone from the start of inflation will encompass the entire portion of the $\tau_{\rm CMB}$ hypersurface visible to us today, at $\tau_0$. If inflation begins even earlier, such that $\Delta \tau_{\rm infl} \ge \tau_{\infty}$, then any two spacetime points within our cosmic event horizon will have past lightcones that intersect at some time since the beginning of inflation. 
%{\bf Remake with $z_A=z_B=50$ instead of 10 to remove coincidence that $\tau_{\rm CMB}\sim |\tau_{\rm AB}|$}
\label{fig:Inflation}
}
\efig
%%%%%%%%%%%%%%%%%%%%%%%%%%%%%%%%%%%%%%%%%%%%%%%%%%%%%%%%%%%%%%%%%

From Eq. (\ref{tauinflation}) we find
%%%%%%%%
\beq
\Delta \tau_{\rm infl} = \tau (t_{\rm end}) - \tau (t_i) = \frac{1}{a_{\rm end} } \left( \frac{H_0}{H_I} \right)\left[ e^{N} - 1 \right] ,
\label{Deltatauinfl2}
\eeq
where $t_i$ is the cosmic time corresponding to the beginning of inflation, $H_I$ is the value of the Hubble constant during inflation, and $e^N = a_{\rm end} / a_i \gg 1$, where $N$ is the total number of efolds during inflation. We may estimate $a_{\rm end}$ by assuming instant reheating to a radiation-dominated phase that persists between $a_{\rm end}$ and $a_{\rm eq} = a (t_{\rm eq})$, where $t_{\rm eq}$ is the time of matter-radiation equality. From Eq. (\ref{aRD}) we have 
%%%%%%%
\beq
a_{\rm end} = a_{\rm eq} \left( \frac{t_{\rm end} }{ t_{\rm eq} } \right)^{1/2} \simeq a_{\rm eq} \left( \frac{ N}{H_I t_{\rm eq} } \right)^{1/2} ,
\label{aend1}
\eeq
upon using $N = H_I ( t_{\rm end} - t_i ) \simeq H_I t_{\rm end}$ during inflation. We also have $a_{\rm eq} / a_0  = 1/ (1 + z_{\rm eq})$. Using our normalization that $a_0 = a (t_0) = 1$, we find
%%%%%%%
\beq
a_{\rm end} \simeq \frac{1}{(1 + z_{\rm eq} )} \left( \frac{ N}{ H_0 t_{\rm eq} } \right)^{1/2} \left( \frac{ H_0}{H_I} \right)^{1/2}
\label{aend2}
\eeq
and therefore Eqs. (\ref{Deltatauinfl1}) and (\ref{Deltatauinfl2}) become
%%%%%%%%%%
\beq
N^{-1/2} \> e^N \geq \frac{1}{( 1 + z_{\rm eq} ) } \left( \frac{1}{ H_0 t_{\rm eq} } \right)^{1/2} \left( \frac{ H_I}{H_0} \right)^{1/2} \left( \tau_0 - 2 \tau_{\rm CMB} \right) .
\label{Nlimit1}
\eeq

Using Eq. (\ref{eq:tau}) with $a_e = a_{\rm CMB} = 1/(1 + z_{\rm CMB})$, we find $\tau_{\rm CMB} = \taucmbnumdim$ and hence $H_0^{-1} \tau_{\rm CMB} = \taucmbnum$ Gyr; putting $a (t_0 ) = 1$ in Eq. (\ref{eq:tau}) yields $\tau_0 = \tauonumdim$ and hence $H_0^{-1} \tau_0  = \tauonum$ Gyr. The latest observations yield $z_{\rm eq} = 3391$ \cite{ade13}, and hence
%%%%%%
\beq
t_{\rm eq} = H_0^{-1} \int_{z_{\rm eq}}^\infty \frac{dz'}{(1 + z') E (z')} = 5.12 \times 10^4 \> {\rm yr} = 1.61 \times 10^{12} \> {\rm sec} .
\label{teq}
\eeq
Recent observational limits on the ratio of primordial tensor to scalar perturbations constrain $H_I \leq 3.7 \times 10^{-5} \> M_{\rm pl}$ \cite{PlanckInflation13}, where $M_{\rm pl} = (8 \pi G)^{-1/2} = 2.43 \times 10^{18}$ GeV is the reduced Planck mass. In ``natural units" (with $c = \hbar = 1$), $1 \>\>\> {\rm GeV}^{-1} = 6.58 \times 10^{-25} \> {\rm sec} = 2.09 \times 10^{-41} \> {\rm Gyr}$, and hence $H_0 = 100 h \Hou = 2.13 h \times 10^{-42} \> {\rm GeV}$, with current best-fit value $h = \hnum$. Eq. (\ref{Nlimit1}) therefore becomes
%%%%%%%%
\beq
N \geq 65.6 .
\label{Nconstraint}
\eeq
Inflation will solve the horizon problem if it persists for at least $N = 65.6$ efolds. 

As is clear from Fig. \ref{fig:Inflation}, if $\Delta \tau_{\rm infl}  \geq \tau_0$, then any two spacetime points within our past lightcone from today will themselves have past lightcones that intersect at some time since the beginning of inflation. Because $\tau_{\rm CMB} \ll \tau_0$, the additional number of efolds of inflation required to satisfy $\Delta \tau_{\rm infl} \geq \tau_0$ rather than Eq. (\ref{Deltatauinfl1}) is $\Delta N = 0.04$, or $N \geq 65.64$. Moreover, if $\Delta \tau_{\rm infl} \geq \tau_\infty$, then any two spacetime points within our entire cosmic event horizon will have past lightcones that intersect at some time since the beginning of inflation. Given $\tau_\infty = 4.33$ (and hence $H_0^{-1} \tau_\infty = 62.90$ Gyr), the additional efolds beyond the limit of Eq. (\ref{Deltatauinfl1}) required to satisfy $\Delta \tau_{\rm infl} \geq \tau_\infty$ is $\Delta N = 0.35$, or a total of $N \geq 65.95$ efolds. Hence virtually any scenario in which early-universe inflation persists long enough to solve the horizon problem will also result in every spacetime point within our cosmic event horizon sharing a common past causal domain.

%----------------------------------------------------------------------------------------------------------------------------------------------------------------------------------

%----------------------------------------------------------------------------------------------------------------------------------------------------------------------------------
\section*{Appendix B. Hubble Sphere and Apparent horizon}
\label{sec:app}
%----------------------------------------------------------------------------------------------------------------------------------------------------------------------------------
We now demonstrate that object pairs in our Universe beyond the causal-independence redshift $z_{\rm ind} > \zindnum$, which have no shared causal pasts since inflation, are also moving away from us at speeds $v_{\rm rec}$ exceeding the speed of light; although objects with current recession velocities $c < v_{\rm rec} \le \vrecindnum c$ will still have a shared causal past with our worldline.  Calculations assume cosmological parameters $\vec{\Omega}$ from Eq~\ref{Omegavalues}.

One might assume that objects would lose causal contact with us and become unobservable if they are currently receding at speeds faster than light. In reality, astronomers today routinely observe light from objects in our universe at redshifts corresponding to superliminal recession velocities (see \cite{davis04,lewisj07}, although see also \cite{bikwa12}).  %Surprisingly, whether we can observe objects that are currently moving away from us superluminally depends subtly on the specific expansion history of a given FLRW cosmology.  
Note that general relativity allows superluminal recession velocities due to cosmic expansion ($v_{\rm rec} = R_0 \dot{a} \chi > c$), though it also requires that objects move with subluminal peculiar velocities ($v_{\rm pec} = R_0 a \dot{\chi} < c$). The so-called ``Hubble sphere" denotes the comoving distance beyond which objects' radial recession velocities exceed the speed of light, $v_{\rm rec} > c$. As $\tau \rightarrow \tau_{\infty}$ the Hubble sphere asymptotes to the cosmic event horizon; see Fig. \ref{fig:Hub}.  

The radial, line-of-sight recession velocity in an FLRW metric is given by
\beqn
\label{eq:vrec}
v_{\rm rec}  =  R_0  \dot{a}  \chi = c  a E(a) \int_{a}^{1} \frac{da'}{a'^2 E(a')}, 
\eeqn
upon using Eq. (\ref{R0}) for $R_0$, Eq. (\ref{Edef}) for $E (a)$, and Eq. (\ref{chiz}) for $\chi$. Eq. (\ref{eq:vrec}) can be used without corrections if the object is at a redshift large enough so that peculiar velocities are negligible compared to cosmic expansion ($a\dot{\chi} \ll \dot{a} \chi$ for $z \gtrsim \vpecznum$ \cite{davis11}). At a given time, $a (t)$, the Hubble sphere is located at a comoving distance $\chi_{\rm hs}$ at which $v_{\rm rec} = c$. Using Eq. (\ref{eq:vrec}) and $R_0 = c / H_0$, the comoving distance $\chi_{\rm hs}$ is given by
\beqn
\label{eq:Rhs}
\chi_{\rm hs}  =  \frac{H_0}{\dot{a}}=\frac{1}{aE(a)} = \int_{a_{\rm hs}}^{1} \frac{da'}{a'^2 E(a')} , 
\eeqn
where $z_{\rm hs} = a_{\rm hs}^{-1} -1$. Note that by our normalization conventions $a (t_0) = 1$ and $E (a (t_0)) = 1$; therefore $\chi_{\rm hs} = 1$, which yields $z_{\rm hs}(t_0)  = \zhsnum$ for $\vec{\Omega}$ as in Eq. (\ref{Omegavalues}). The current Hubble sphere redshift $z_{\rm hs}  = \zhsnum$ is thus less than the current causal-independence redshift, $z_{\rm ind} = \zindnum$.  Using parameters $\vec{\Omega}$ in Eq. (\ref{Omegavalues}), we find that objects at $z =\zindnum$ have recession velocities of $v_{\rm rec} = \vrecindnum  c$, so objects that are currently receding from us faster than light in the range $c < v_{\rm rec} \leq \vrecindnum  c$ still have a shared causal past with our worldline since $\tau > 0$.

Another quantity of interest is the apparent horizon \cite{faraoni11,bousso11} or the minimally anti-trapped hypersurface \cite{starkman99}, which is located at a line-of-sight comoving distance $\chi_{\rm ah}$ given by
\beqn
\label{eq:Rah}
\chi_{\rm ah} = \frac{1}{\sqrt{ ( \dot{a} / H_0 )^2 - \Omega_k } } = \frac{1}{\sqrt{ \left[a E(a)\right]^2 - \Omega_k } } = \frac{1}{ \sqrt{ \Omega_\Lambda a^{2} + \Omega_M a^{-1} + \Omega_R a^{-2} } } .
\eeqn
Hence $\chi_{\rm ah} = \chi_{\rm hs}$ when $\Omega_k = 0$ (also see \cite{faraoni11}). In our flat universe, the redshifts of the apparent horizon and the Hubble sphere are thus identical, and since $z_{\rm ind} > z_{\rm hs}$, objects that have no shared causal past with our worldline since the big bang, with redshifts $z > \zindnum > \zhsnum$, are also by necessity moving superluminally.

%----------------------------------------------------------------------------------------------------------------------------------------------------------------------------------

%--------------------------------------------------------------------------------------------------------

\acknowledgments{
It is a pleasure to thank Alan Guth for helpful discussions. Bruce Bassett provided useful comments on an early draft. This work was supported in part by the U.S. Department of Energy (DoE) under contract No. DE-FG02-05ER41360. ASF was also supported by the U.S. National Science Foundation (NSF) under grant SES 1056580. The authors made use of the MILLIQUAS - Million Quasars Catalog, Version 3.1 (22 October 2012), maintained by Eric Flesh (\texttt{http://heasarc.gsfc.nasa.gov/W3Browse/all/flesch12.html}).
}

%----------------------------------------------------------------------------------------------------------------------------------------------------------------------------------
%\begin{thebibliography}{999}
%@CONTROL{apsrev41Control,author="3",editor="1",pages="0",title="1",year="1"}

%\bibliographystyle{apsrev4-1}  %  BibTeX styles for use for Phys. Rev. journals, References listed in order of placement within the text
%\bibliographystyle{apsrmp4-1} %  BibTeX styles for use for Rev. Mod. Phys.
%\bibliographystyle{aipauth4-1} %  BibTeX styles for AIP journals with author/year style citations
%\bibliographystyle{aipnum4-1} %  BibTeX styles for AIP journals with numerical style citations

%Don't use \bibliographystyle{} if you want longbibliography in \documentclass[12pt,aps,showpacs,notitlepage,longbibliography]{revtex4-1} to work where article titles are listed in the bibilography as well

\bibliography{./selection,./sn,./sngroup} % *.bib files in working directory
%\bibliography{../bib/selection,../bib/sn,../bib/sngroup}

%----------------------------------------------------------------------------------------------------------------------------------------------------------------------------------

%----------------------------------------------------------------------------------------------------------------------------------------------------------------------------------
\end{document}